\documentclass[aps,prl,onecolumn,showpacs,showkeys,preprintnumbers]{revtex4}
\usepackage{eurosym}
\usepackage{amsmath}
\usepackage{dcolumn}
\usepackage{bm}
\usepackage{subfigure}
\usepackage{amsfonts}
\usepackage{amssymb}
\usepackage{makeidx}
\usepackage{epsfig}
\usepackage{graphicx}

\begin{document}

\title{\textbf{Generalized Lagrangian Path approach to manifestly-covariant
quantum gravity theory}}
\author{Massimo Tessarotto}
\affiliation{Department of Mathematics and Geosciences, University of Trieste, Via
Valerio 12, 34127 Trieste, Italy}
\affiliation{Institute of Physics, Faculty of Philosophy and Science, Silesian University
in Opava, Bezru\v{c}ovo n\'{a}m.13, CZ-74601 Opava, Czech Republic}
\author{Claudio Cremaschini}
\affiliation{Institute of Physics and Research Center for Theoretical Physics and
Astrophysics, Faculty of Philosophy and Science, Silesian University in
Opava, Bezru\v{c}ovo n\'{a}m.13, CZ-74601 Opava, Czech Republic}
\date{\today }

\begin{abstract}
A trajectory-based representation for the quantum theory of the
gravitational field is formulated. This is achieved in terms of a
covariant Generalized Lagrangian-Path (GLP) approach which relies
on a suitable statistical representation of Bohmian Lagrangian
trajectories, referred to here as \emph{GLP-representation.} The
result is established in the framework of the manifestly-covariant
quantum gravity theory (CQG-theory) proposed recently and the
related CQG-wave equation advancing in proper-time the quantum
state associated with massive gravitons. Generally non-stationary
analytical solutions for the CQG-wave equation with non-vanishing
cosmological constant are determined in such a framework, which
exhibit Gaussian-like probability densities that are
non-dispersive in proper-time. As a remarkable outcome of the
theory achieved by implementing these analytical solutions, the
existence of an emergent gravity phenomenon is proved to hold.
Accordingly, it is shown that a mean-field background space-time
metric tensor can be expressed in terms of a suitable statistical
average of stochastic fluctuations of the quantum gravitational
field whose quantum-wave dynamics is described by GLP
trajectories.
\end{abstract}

\maketitle

\section{1 - Introduction}

The search of a theory of quantum gravity which is consistent both with the
principles of quantum mechanics \cite{Messiah} as well as with the
postulates of the classical Einstein theory of General Relativity (GR) \cite%
{ein1,LL,gravi} has represented so far one of the most challenging and
hard-to-solve conceptual problems of mathematical and theoretical physics\
alike. The crucial issue is about the possibility of achieving in the
context of either classical or quantum relativistic theories, and in
particular for a quantum theory of gravity, a truly coordinate- (i.e.,
frame-) independent representation, namely which satisfies - besides the
general covariance principle - also the so-called principle of manifest
covariance.\textbf{\ }In fact, although the choice of special coordinate
systems is always legitimate for all physical systems either discrete or
continuous, including in particular classical and quantum gravity, the
intrinsic objective nature of physical laws makes them frame-independent.

However, in order that these principles can actually apply, a background
space-time picture must hold. This means, more precisely, that a suitable
classical curved space-time\ $\left\{ \mathbf{Q}^{4},\widehat{g}\right\} $\
must exist with respect to which both general covariance principle and
principle of manifest covariance can be prescribed.\textbf{\ }As a
consequence, when parametrized with respect to a coordinate system $r\equiv
\left\{ r^{\mu }\right\} $ the same space-time must be\ endowed with a
well-defined (i.e., uniquely prescribed and hence deterministic) symmetric
metric tensor $\widehat{g},$ represented equivalently in terms of its
covariant $\widehat{g}\equiv \left\{ \widehat{g}_{\mu \nu }\right\} $ and
countervariant $\widehat{g}\equiv \left\{ \widehat{g}^{\mu \nu }\right\} $
forms , which is referred to in the following as the "\textit{background}"
field tensor.\textbf{\ }In particular, $\mathbf{Q}^{4}$\ can be identified
with a time-oriented $4-$dimensional Riemann space-time. Thus, although the
precise choice of the same background space-time itself remains in principle
arbitrary, as a consequence of the principle of manifest covariance it
should always be possible to represent all quantum observables (of the
theory), including the corresponding quantum Hamiltonian operator and
quantum canonical variables/operators (see below), in $4-$tensor form. This
requires to cast them exclusively as $4-$tensor fields with respect to the
group of local point transformations (LPT group)%
\begin{equation}
r\equiv \left\{ r^{\mu }\right\} \rightarrow r^{\prime }\equiv \left\{
r^{\prime \mu }\right\} =r^{\prime }(r)  \label{LPT}
\end{equation}%
mapping $\left\{ \mathbf{Q}^{4},\widehat{g}\right\} $\ in itself \cite{noi4}.

In such\ a framework $\widehat{g}$\ is considered as a classical (i.e.,
deterministic) tensor field, to be identified as the metric tensor field of $%
\left\{ \mathbf{Q}^{4},\widehat{g}\right\} $ which - as such - determines
the geometric properties of the same space-time. This means more precisely
that:

\emph{Prescription a - }Its covariant and countervariant components, i.e.,
respectively $\widehat{g}_{\mu \nu }$ and $\widehat{g}^{_{\mu \nu }}$, must
lower and raise tensor indices of arbitrary tensor fields and also prescribe
the standard connections (Christoffel symbols) appearing in the covariant
derivatives.

\emph{Prescription b - }It determines the Ricci tensor, the Ricci $4-$scalar
and the coupling contained in the stress-energy tensor due to external
sources, in the sequel respectively identified with the symbols $\widehat{R}%
_{\mu \nu }$ $\equiv R_{\mu \nu }(\widehat{g}),$ $\widehat{R}$ $\equiv R(%
\widehat{g})\equiv \widehat{g}^{\alpha \beta }\widehat{R}_{\alpha \beta }$
and $\widehat{T}_{\mu \nu }=T_{\mu \nu }(\widehat{g})$.

\emph{Prescription c - }As a consequence, $\widehat{g}$ can be identified
with a particular solution of\ the Einstein field equations%
\begin{equation}
\widehat{R}_{\mu \nu }-\frac{1}{2}\left[ \widehat{R}-2\Lambda \right]
\widehat{g}_{\mu \nu }=\frac{8\pi G}{c^{4}}\widehat{T}_{\mu \nu },
\label{EINSTEIN FIELD EQS}
\end{equation}%
where as usual\textbf{\ }$\Lambda $\textbf{\ }denotes the cosmological
constant.

\emph{Prescription d - }$\widehat{g}$ determines uniquely \emph{the Riemann
distance} $s$, or \emph{proper-time}, on the space-time $\left\{ \mathbf{Q}%
^{4},\widehat{g}\right\} $ by means of the $4-$scalar equation
\begin{equation}
ds^{2}=\widehat{g}_{\mu \nu }(r,s)dr^{\mu }dr^{\nu }.  \label{ds^2}
\end{equation}%
One notices that, in accordance with Ref.\cite{noi4}, here $dr^{\mu }\equiv
dr^{\mu }(s)$\ and $ds$\ identify respectively the $4-$tensor displacement
and its corresponding $4-$scalar line-element (arc length), both evaluated
along a suitable worldline. For this purpose the latter is identified with
an arbitrary geodetics $r(s)\equiv \left\{ r^{\mu }(s)\right\} $\ belonging
to $\left\{ \mathbf{Q}^{4},\widehat{g}\right\} $ that crosses an arbitrary $%
4-$position $r^{\mu }\equiv r_{o}^{\mu }$, and hence fulfills the initial
condition $r^{\mu }(s_{o})=r_{o}^{\mu }$, at some proper time $s_{o}$\
(which for definiteness can always be set $s_{o}=0$). As a consequence, the
definition of proper time remains unambiguous and unique also for arbitrary
finite values of $s\in I$ (with $I\equiv
\mathbb{R}
$ the real axis), being identified with the arc length along the (unique)
geodetics $r(s)\equiv \left\{ r^{\mu }(s)\right\} $\ joining $r^{\mu
}(s_{o})=r_{o}^{\mu }$\ with an arbitrary $4-$position $r_{1}^{\mu },$\
i.e., such that $r^{\mu }(s_{1})=r_{1}^{\mu }$ for a given $s_{1}$\ assumed
to exist. For example, the proper time can always be defined along an
appropriate observer geodetics.

\emph{Prescription e - }One notices that in principle the background metric
tensor might be taken of the form $\widehat{g}(r,s)\equiv \left\{ \widehat{g}%
_{\mu \nu }(r,s)\right\} $, i.e., allowed to depend explicitly also on the
proper time $s.$\ In the following, however, we shall restrict the treatment
to the customary case in which the background metric tensor solution of the
Einstein field equations is purely dependent only on the $4-$position $%
r^{\mu }$, namely is of the form%
\begin{equation}
\widehat{g}=\widehat{g}\left( r\right) ,  \label{stationary metric tensor}
\end{equation}%
which identifies a\textbf{\ }\textit{stationary metric tensor}.

Next, let us consider the prescription holding for the Lagrangian continuum
coordinates $g\equiv \left\{ g^{\mu \nu }\right\} $\ and the conjugate
momentum operator $\pi \equiv \left\{ \pi ^{\mu \nu }\right\} $, again both
to be considered as $4-$tensor fields with respect to the group of local
point transformations (\ref{LPT}):

\emph{Prescription f -} As a consequence of the stationarity assumption (\ref%
{stationary metric tensor}), for all sets $(r,s)\in \left\{ \mathbf{Q}^{4},%
\widehat{g}\right\} \times I$\ tensor decompositions of the form%
\begin{equation}
\left\{
\begin{array}{c}
g(r,s)=\widehat{g}(r)+\delta g(r,s), \\
\pi (r,s)=\delta \pi (r,s),%
\end{array}%
\right.  \label{TENSOR DECOMPOSITION}
\end{equation}%
will be assumed to hold for the quantum gravity theory, with $\delta
g(r,s)\equiv \left\{ \delta g_{\mu \nu }(r,s)\right\} $\ and\ $\delta \pi
(r,s)\equiv \left\{ \delta \pi _{\mu \nu }(r,s)\right\} $\ denoting the
corresponding \emph{quantum fluctuations}, represented by a coordinate
displacement field and momentum operator which by assumption may depend
explicitly on the variables $(r,s)$.

A promising new scenario for quantum gravity fulfilling these requirements
has recently been established in Refs.\cite{noi0,noi1,noi2,noi5,noi6,noi7}.
This is realized by the theory of manifestly-covariant quantum gravity,
denoted as CQG-theory, which is based on the manifestly-covariant canonical
quantization ($g-$\emph{quantization})\emph{\ }of the classical Hamiltonian
state $\left\{ g(r,s),\pi (r,s)\right\} $. It must be clarified that in the
present treatment the concept of manifest covariance means that CQG-theory
is realized by a formulation in which all classical and quantum Hamiltonian
field variables or operators, including continuum coordinates, conjugate
momenta and Hamiltonian densities transform as $4-$tensors, i.e., fulfill
covariance tensor transformation laws with respect to the group of local
point transformations (\ref{LPT}). Although a manifestly-covariant theory of
this type need not necessarily to be unique, the involved notion of manifest
covariance given here is certainly unambiguously determined when the
background space-time $\left\{ \mathbf{Q}^{4},\widehat{g}\right\} $ is
prescribed. On the other hand, an alternative route is also available. This
is based on the preliminary introduction of a non-canonical mapping in which
the classical (and hence also the quantum) Hamiltonian state $\left\{
g(r,s),\pi (r,s)\right\} $\ is mapped by means of a diffeomorphism onto a
suitable set of non-canonical variables%
\begin{equation}
\left\{ g(r,s),\pi (r,s)\right\} \Leftrightarrow \left\{ \eta (r,s),\chi
(r,s)\right\} ,
\end{equation}%
in which, however, $\eta (r,s)\equiv \left\{ \eta _{\alpha \beta
}(r,s)\right\} $\ and $\chi (r,s)\equiv \left\{ \chi _{\alpha \beta
}(r,s)\right\} $\ are not represented by $4-$tensor variables.\ When
expressed in terms of the transformed variables $\left\{ \eta (r,s),\chi
(r,s)\right\} $\ CQG-theory does not lose obviously the property of
covariance (its equations remain covariant with respect to the LPT-group)
although its variables (i.e., $\left\{ \eta (r,s),\chi (r,s)\right\} $) are
not represented by $4-$tensors. Such a notion will be referred to as
property of plain covariance of the theory. The distinction between the two
notions of covariance (manifest or plain) is, however, important. In fact
manifest covariance represents a stronger condition for the realization of a
quantum theory of gravitational field with respect to literature approaches
which, instead, may or may not rely on weaker notions of covariance such as
that of plain covariance (see also subsequent discussion in Section 2).

As such, CQG-theory is endowed with a number of key features, since: A) it
preserves the background metric tensor $\widehat{g}(r)$\ which is identified
with a classical field tensor;\textbf{\ }B) it satisfies the \emph{quantum
unitarity principle,} i.e., the quantum probability is conserved; C) it is
\emph{constraint-free}, in the sense that the quantum Lagrangian variables $%
g\equiv g(r,s)$\ are identified with independent tensor fields; D) it is
\emph{non-perturbative} so that the quantum fluctuations $\delta g(r,s)$\
and $\delta \pi (r,s)$\ need not be regarded as asymptotically "small" in
some appropriate sense with respect to the background metric tensor $%
\widehat{g}(r)$. Its foundations (for a detailed discussion see Ref.\cite%
{noi5}) lie on the preliminary establishment of a variational formulation of
GR achieved in the context of a\ covariant DeDonder-Weyl-type approach to
continuum field-Hamiltonian dynamics \cite%
{donder,weyl,sym3,sym4,sym5,sym7,sym8,sym9} in which the background
space-time $\left\{ \mathbf{Q}^{4},\widehat{g}\right\} $ is considered
prescribed \cite{noi1,noi2}.

In the following we intend to shed further light on key aspects of the
CQG-theory which are intimately related with its consistent realization.
These include in particular two crucial "\textit{tests of consistency}" for
CQG-theory which should actually be regarded as mandatory physical
prerequisites for any quantum theory of gravity fulfilling both the
principles of general and manifest covariance.

The first one is that,\ although quantum corrections may in principle occur
\cite{noi1,noi2}, it must be possible to preserve the functional form of the
Einstein field equations consistent with the\emph{\ }so-called\emph{\
emergent gravity picture}. More precisely, the latter equations should
follow uniquely from quantum theory itself \emph{without performing the\
semiclassical continuum limit}\ (namely obtained letting in particular $%
\hslash \rightarrow 0$; see for example Ref.\cite{Han} where the derivation
of the Einstein field equation was discussed in the context of loop quantum
gravity).\textbf{\ }This property will be referred to here as\textbf{\ }"%
\emph{first-type emergent-gravity paradigm}".

The second test of consistency, to be investigated here,\ refers instead to
the validity of an emergent-gravity picture also for the deterministic
background metric tensor $\widehat{g}(r),$ in the sense that the same $%
\widehat{g}(r)$\ should be prescribed by means of a suitably-defined
quantum/stochastic expectation value of the quantum state.\textbf{\ }This
property will be denoted here as "\emph{second-type emergent-gravity\
paradigm}".\textbf{\ }A basic requirement needed for its verification is the
determination of a suitable class of particular solutions of the quantum
wave-function, i.e., the CQG-wave equation for the quantum state $\psi
(g,r,s)$\ earlier pointed out in Ref.\cite{noi6}.

With these tasks in mind, in the following Eulerian and Lagrangian
representations are preliminarily distinguished for the CQG-wave equation\
and its corresponding set of quantum hydrodynamic equations (QHE). The
latter are implied by the Madelung representation \cite{Madelung1928} of the
quantum wave function written in Eulerian form $\psi \equiv \psi \left(
g,r,s\right) $, namely distinguishing the dependences in terms of the
Lagrangian coordinates $g\equiv \left\{ g_{\mu \nu }\right\} $ and the
parameters $(r,s)$ as%
\begin{equation}
\psi (g,r,s)=\sqrt{\rho (g,r,s)}\exp \left\{ \frac{i}{\hslash }%
S^{(q)}(g,r,s)\right\} .  \label{Axiom-5-1}
\end{equation}%
Here the real fields $\left\{ \rho ,S^{(q)}\right\} \equiv \left\{ \rho
(g,r,s)=\left\vert \psi (g,r,s)\right\vert ^{2},S^{(q)}(g,r,s)\right\} $\
identify the quantum fluid $4-$scalar fields written in Eulerian form,
namely the quantum probability density function (PDF) and the quantum
phase-function. In particular, the intent of the investigation concerns the
introduction of a trajectory-based or Lagrangian representation of
CQG-theory (see subsequent Sections 4 and 5), to be distinguished from the
Eulerian one (see Section 3) and referred to here as \emph{Generalized
Lagrangian-path approach} to CQG-theory. This goal is obtained by means of
an appropriate parametrization of the corresponding set of quantum
hydrodynamic equations, following in turn from the CQG wave-equation and
based on the Madelung representation recalled above (see Eq.(\ref{Axiom-5-1}%
)). More precisely, this concerns the investigation of:

\begin{itemize}
\item \emph{Goal \#1}: Explicit solutions of the CQG-quantum hydrodynamic
equations satisfying suitable physical requirements.

\item \emph{Goal \#2}:\emph{\ }The "emergent" character of the classical
background space-time metric tensor\textbf{\ }$\widehat{g}(r)$,\ to be
determined in terms of quantum theory. Accordingly, the background metric
tensor $\widehat{g}(r)$\ should be identified with a suitably-defined
quantum expectation value of the quantum state, i.e., weighted in terms of
the corresponding quantum probability density (PDF).

\item \emph{Goal \#3}: The existence of\ either stationary or, more
generally, non-stationary solutions with respect to the proper-time $s$,
i.e., explicitly dependent on $s$, for the quantum state $\psi $ expressed
via the Madelung representation (see Eq.(\ref{Axiom-5-1})).

\item \emph{Goal \#4}:\emph{\ }The search of Gaussian-like or Gaussian
realizations for the quantum PDF $\rho $.

\item \emph{Goal \#5}:\emph{\ }The search of separable solutions of the
quantum Hamilton-Jacobi (H-J) equation in terms of the quantum
phase-function $S^{(q)}$ and the investigation of their qualitative
properties and in particular their asymptotic behavior for $s\rightarrow
+\infty $.
\end{itemize}

For the tasks indicated above, in close similarity with non-relativistic
quantum mechanics (see Refs.\cite{Tessarotto2016,Tessarotto2016-b}), two
choices are in principle available. The first one is based on the
introduction of deterministic Lagrangian trajectories $\left\{ g(s),s\in
I\right\} $, or Lagrangian-Paths (LP), analogous to those adopted in the
context of the Bohmian representation of non-relativistic quantum mechanics%
\textbf{\ }\cite{Nelson,new1,Holland1,Poirier,Holland2,Poirier2,Schi,Parlant}%
. This provides a Bohmian interpretation (of CQG-theory) which is
ontologically equivalent to CQG-theory itself \cite{boom1}. Hence, the
tensor field $\delta g(s)\equiv \delta g_{L}(s)$\ is uniquely determined by
means of a map of the type%
\begin{equation}
s\rightarrow \delta g_{L}(s)\equiv \delta g_{L}(r(s),s),  \label{LP}
\end{equation}%
with $r=r(s)$\ denoting the parametrization in terms of geodetic curves
associated with the classical background field tensor $\widehat{g}(r)\equiv
\widehat{g}(r(s))$ (see \textit{Prescription d} above,\textbf{\ }Ref.\cite%
{noi5} and related discussion in Section 4 below), so that in terms of $%
g_{L}(s)\equiv g(s)$ it follows that $\left\{ g(s),s\in I\right\} \equiv
\left\{ g_{L}(s)=\widehat{g}(r(s))+\delta g_{L}(s),s\in I\right\} $. The
second choice, instead, and the one which is at the basis of the GLP
trajectory-based approach (or \emph{GLP-representation}) adopted here, is
achieved in terms of suitable stochastic, i.e., intrinsically non-unique,
Lagrangian trajectories which are referred to here as \emph{Generalized
Lagrangian Paths} (GLP). Such a notion, which is inspired and extends to
CQG-theory the analogous approach earlier developed for non-relativistic
quantum mechanics \cite{Tessarotto2016}, is based on a suitable
generalization of the concept of LP (see Section 5 below).\textbf{\ }In such
a context each deterministic LP $\left\{ g(s),s\in I\right\} $ is replaced
with a continuous statistical ensemble of \emph{stochastic GLP\ trajectories}
$\left\{ G(s),s\in I\right\} $. More precisely, introducing in analogy with
Eq.(\ref{TENSOR DECOMPOSITION}) the tensor decomposition%
\begin{equation}
G(s)=\widehat{g}(r(s))+\delta G(s),  \label{TENSOR DECOMPOSITION 2}
\end{equation}%
with $\delta G(s)\equiv \left\{ \delta G_{\mu \nu }(r(s),s)\right\} $\ being
a suitable tensor field denoted as\textbf{\ }\emph{GLP-displacement}\textbf{%
\ }to be later defined, each GLP trajectory\textbf{\ }%
\begin{equation}
\left\{ G(s),s\in I\right\} \equiv \left\{ \widehat{g}(r(s))+\delta
G(s),s\in I\right\}   \label{GLP definition}
\end{equation}%
is parametrized in terms of the displacement field, to be considered as a
stochastic field tensor,%
\begin{equation}
\Delta g=\delta g(s)-\delta G(s),  \label{SETTING}
\end{equation}%
with $\Delta g\equiv \left\{ \Delta g_{\mu \nu }\right\} $\ denoting a
suitable constant second-order tensor field referred to here as \emph{%
stochastic displacement field tensor}. For definiteness, it is required that
its covariant components at proper-times $s$\ and $s_{o}$, $\Delta g_{\mu
\nu }(s)=\delta g_{\mu \nu }(s)-\delta G_{\mu \nu }(s)$\ and $\Delta g_{\mu
\nu }(s_{o})=\delta g_{\mu \nu }(s_{o})-\delta G_{\mu \nu }(s_{o})$, are
prescribed so that for all $s,s_{o}\in I$%
\begin{equation}
\Delta g_{\mu \nu }(s)=\Delta g_{\mu \nu }(s_{o}).  \label{PREVIOUS}
\end{equation}%
Then, this implies that its counter-variant components $\Delta g^{\mu \nu
}(s)$\ and $\Delta g^{\mu \nu }(s_{o})$ can be equivalently determined in
terms of the prescribed field tensors $\widehat{g}_{\mu \nu }(r)\equiv
\widehat{g}(r(s))$\ or\ $\widehat{g}_{\mu \nu }(r_{o})\equiv \widehat{g}%
(r(s_{o}))$, so that one has necessarily for all $s,s_{o}\in I$\ that\
\begin{equation}
\Delta g^{\mu \nu }(s)=\Delta g^{\mu \nu }(s_{o})
\label{CONSTANCY CONDITION-2}
\end{equation}%
too. As a consequence, each GLP trajectory is actually represented by a
configuration-space curve of the type\textbf{\ }$\left\{ G(s),s\in I\right\}
\equiv \left\{ \widehat{g}(r(s))+\delta g(s)-\Delta g,s\in I\right\} ,$\ so
that upon varying the stochastic displacement field tensor $\Delta g$\ it
actually defines a statistical ensemble of trajectories. In terms of them,
i.e., by parametrizing the CQG wave-function $\psi (g,r,s)$\ (or
equivalently the corresponding quantum fluid fields) in terms of the
GLP-displacement $\delta G(s)=\delta g(s)-\Delta g,$\ the\ \emph{%
GLP-representation} of CQG-theory is then achieved. This amounts to
introduce the composed mapping $\psi (g,r,s)\rightarrow \psi (G(s),\Delta g,%
\widehat{g},r,s)$, where $\psi (G(s),\Delta g,\widehat{g},r,s)$ denotes the
\emph{GLP-parametrized quantum wave-function} in which the dependence in
terms of the displacement tensor field $\Delta g$\ is explicitly allowed.

As shown in Section 5, the adoption of the GLP parametrization for
CQG-theory actually leaves unchanged the underlying axioms established in
Ref.\cite{noi6}, thus providing a Lagrangian representation of CQG-theory
which is ontologically equivalent to CQG-theory itself. The remarkable new
aspects of the GLP\ formalism, however, are that it will be shown: a) first,
to determine a solution method for the CGQ-wave equation, to be referred to
here as \emph{GLP-approach}, permitting the explicit construction of
physically-relevant particular realizations of the CGQ-quantum state $\psi
\left( s\right) $; b) second, to realize quantum solutions which are
consistent with the emergent-gravity picture. In particular, for this
purpose, the background field tensor will be shown to be determined
equivalently either in terms of quantum expectation values or via a
suitably-prescribed stochastic average of the quantum field tensor $g_{\mu
\nu }$. This includes the determination of particular solutions of the
CQG-wave equation\ which, consistent with Goal \#1, satisfy the following
physical requirements:\newline

\begin{itemize}
\item \emph{Requirement \#1: }the quantum wave-function $\psi (s)$\ is \emph{%
dynamically consistent}, namely for which the PDF $\rho (g,r,s)\equiv
\left\vert \psi (g,r,s)\right\vert ^{2}$\ associated with the quantum
wave-function $\psi (g,r,s)$\ is globally prescribed and summable in the
quantum configuration space $U_{g}$ in such a way that the corresponding
probability $\left\vert \psi \right\vert ^{2}d(g)$ is similarly globally
conserved for arbitrary subsets of the quantum configuration space $U_{g}$.
As discussed below a prerequisite for meeting such a requirement is the
validity of suitable Heisenberg inequalities earlier determined in Ref.\cite%
{noi7}.

\item \emph{Requirement \#2:} $\psi (s)$\ exhibits the\ \emph{explicit
dependence in terms of a stochastic observable,} so to yield a so-called
Stochastic-Variable Approach to quantum theory \cite%
{Tessarotto2016,boom2,boom3,boom4}. In the context of CQG-theory this should
be generally identified with a $4-$tensor field depending on the physical
quantum observable $g_{\mu \nu }(r,s)$ and realizing a stochastic variable
endowed with a stochastic probability density, i.e., dependent on a suitable
stochastic field. Such a stochastic field will be identified in the
following with the second-order real and observable stochastic displacement
field tensor $\Delta g=\left\{ \Delta g_{\mu \nu }\right\} $ defined by Eq.(%
\ref{SETTING})\textbf{\ }which by assumption depends functionally on $g_{\mu
\nu }$ (and hence $\delta g_{\mu \nu }(r,s)$\ too).\newline

\item \emph{Requirement \#3:} the PDF $\rho $\ is endowed with a\emph{\
Gaussian-like }behavior and is\emph{\ non-dispersive }in character, namely
in the sense of assuming that\ in the subset of the proper-time axis $I$ in
which $\psi $ is defined, its probability density $\left\vert \psi
\right\vert ^{2}$\textbf{\ }can be identified for all $s\in I\equiv
\mathbb{R}
$ with a Gaussian-like PDF depending on $\Delta g$\ and $\widehat{g}$, and
thus by itself realizes a stochastic function. These particular solutions of
the CQG-wave equation are generally non-stationary and are required\ to
preserve their Gaussian-like character,\ and therefore to be non-dispersive,
i.e., free of any spreading behavior during the proper-time quantum
dynamical evolution.

\item \emph{Requirement \#4:}\textbf{\ }the quantum wave function holds for
arbitrary realizations of the deterministic background metric tensor $%
\widehat{g}(r)$\textbf{\ }and in particular in the case of vacuum solutions
of the Einstein field equations.
\end{itemize}

Requirements \#1-\#4 are physically motivated.\ More precisely, the first
one is needed to warrant the validity of the quantum unitarity principle,
i.e., the conservation of quantum probability.\textbf{\ }The second
requirement, instead, is instrumental for the present theory. In fact, as
clarified below, the existence of the stochastic tensor observable $\Delta
g(g)$\ is mandatory for the development of a GLP-approach in the context of\
CQG-theory. The third requirement is related to the issue about the physical
origin of the cosmological constant \cite{wein}. The existence of
Gaussian-like solutions for the quantum PDF $\rho \left( s\right) $\ is
mandatory in order to establish the connection between the CQG-theory and
the Einstein field equations and to identify its precise quantum origin in
terms of the Bohm vacuum interaction \cite{bohm1,bohm2,bohm3}. Finally,
requirement \#4 is intimately related to the principle of manifest
covariance and the deterministic character of the background metric tensor $%
\widehat{g}(r)$.

As a further remark, one notices that Requirements \#2 and \#3 are
qualitatively similar to those set at the basis of the GLP-approach
developed for non-relativistic quantum mechanics. These led to the
identification and proof of existence of non-dispersive Gaussian-like, or
even properly Gaussian, particular solutions\ of the Schroedinger equation
originally conjectured by Schr\"{o}dinger himself in 1926 \cite%
{Schroedinger1928}. It is therefore natural to conjecture that analogous
properties should hold in the context of the CQG-theory. As a remarkable
conceptual outcome of the GLP theory, it is then shown that the discovery of
analytical solutions satisfying physical Requirements \#1-\#4 allows for the
investigation of theoretical aspects of the quantization of the
gravitational field which go beyond the framework of so-called
first-quantization, toward inclusion of second-quantization effects. This
refers to quantum interactions of the gravitational field with itself which
are intrinsically proper-time dependent contributions generated by the
quantum wave dynamics retained in the solution of the same background metric
tensor. In particular, in this work the existence of an emergent gravity
phenomenon is displayed, which establishes a precise relationship between
the background metric tensor $\widehat{g}_{\mu \nu }$\ and the quantum field
$g_{\mu \nu }$. In detail, it is shown that $\widehat{g}_{\mu \nu }$ can be
represented as a mean-field background space-time metric tensor provided by
a statistical moment of the Gaussian (or more generally Gaussian-like) PDF $%
\rho $. Hence, from the physical point of view $\widehat{g}_{\mu \nu }$\ can
be effectively interpreted as arising from\ a statistical average of
stochastic fluctuations of the quantum gravitational field $g_{\mu \nu }$\
whose quantum-wave dynamics is described by GLP trajectories.

In detail the structure of the paper is as follows. First, a qualitative
comparison between CQG-theory and literature approaches to quantum gravity
and its Bohmian formulation is proposed in Section 2. The Eulerian
representation of CQG-theory is then presented in Section 3. Subsequently,
the Lagrangian-path and Generalized Lagrangian-path representations are
pointed out in Sections 4 and 5, together with their Bohmian, i.e.,
deterministic, and correspondingly stochastic interpretations. Next,
consistent with the axioms of CQG-theory, in Section 6 the establishment of
the stochastic probability density attached with the stochastic displacement
field tensor $\Delta g$ is achieved.\ This is shown to be necessarily
identified with the initial quantum PDF. In connection with such a
prescription, in the same section the problem is posed of the construction
of generalized Gaussian particular solutions for the quantum PDF $\rho
(g,r,s)$.\ Subsequently, in Section 7 the search of separable solutions of
the corresponding quantum H-J equation is investigated. As a result,
asymptotic conditions are investigated warranting the quantum phase function
to be expressed in terms of polynomials of $\Delta g$. Finally, in Section 8
the main conclusions of the paper are drawn, while Appendices A and B
contain mathematical details of the calculations.

\section{2 - Quantum gravity theories and Bohmian formulation in literature}

This section is intended to provide a summary of the relevant conceptual
features of CQG theory, together with an exhaustive discussion of literature
works dealing with quantum gravity theories and corresponding Bohmian
formulations. The aim of such a comparison with previous literature is
twofold. From one side, we intend pointing out the main differences and
significant progresses of CQG-theory from alternative approaches to quantum
gravity. From the other side, we are interested in stating which are the
common aspects of the present approach with other quantum theories of the
gravitational field, and in which sense CQG-theory and the literature
formulations discussed here can be reconciled or regarded as complementary.
A review of the mathematical foundations of CQG theory and its Hamiltonian
structure is treated separately in Section 3.

We start by noting that according to Ref.\cite{hh1} quantization methods,
both in quantum mechanics and quantum gravity, can be classified in two
classes, denoted respectively as the canonical and the covariant approaches.
These differ in the way in which both the quantum state and the space-time\
are treated. In fact, the canonical quantization approach is based, first on
the preliminary introduction of ($3+1$)$-$ or ($2+2$)$-$decompositions (or
foliations \cite{zzz2,Vaca5,Vaca6})\textbf{\ }for the representation of the
space-time and, second, on the adoption of a quantum state represented in
terms of non-4-tensor continuum fields. As such, by construction these
theories are not covariant with respect to the LPT-group (\ref{LPT}).
Nevertheless they still may retain well-definite covariance properties with
respect to appropriate subgroups of local point transformations. For
example, in the case of the ($3+1$)-decomposition covariance is warranted
with respect to arbitrary point transformations which preserve the same
foliation. In the covariant approaches, instead, typically all physical
quantities including the quantum state are represented exclusively by means
of $4-$tensor fields. so that the property of manifest covariance remains
fulfilled. As a consequence, for these approaches covariant quantization
involves the assumption of some sort of classical background space-time
structure on which a quantum gravity theory is constructed, for example
identified with the flat Minkowski space-time. In order to realize such a
strategy, however, it turns out that the quantum state is typically
represented in terms of superabundant variables. As a consequence in such
cases covariant quantization may also require the treatment of suitable
constraint conditions.

Let us briefly analyze both approaches in more detail, considering first the
canonical approach. A choice of this type is exemplified by the one adopted
by Dirac and based on the Dirac constrained dynamics \cite%
{dirac4,dirac3,cast1,cast2,cast3}. Dirac Hamiltonian approach to quantum
gravity is not manifestly covariant, in reference both to transformation
properties with respect to local as well as non-local point transformations
(see discussion in Ref.\cite{noi4}). In this picture in fact the field
variable is identified with the metric tensor $g_{\mu \nu }$, but the
corresponding "generalized velocity" is defined as $g_{\mu \nu ,0}$, namely
with respect to the \textquotedblleft time\textquotedblright\ component of
the $4-$position. This choice necessarily violates the principle of manifest
covariance \cite{noi1,noi2}. Consequently, in Dirac's canonical theory, the
canonical momentum remains identified with the manifestly non-tensorial
quantity $\pi _{Dirac}^{\mu \nu }=\frac{\partial L_{EH}}{\partial g_{\mu \nu
,0}}$, where $L_{EH}$ is the Einstein-Hilbert variational Lagrangian density.

The same kind of ingredients is at the basis of the approach developed by
Arnowitt, Deser and Misner (ADM theory, 1959-1962 \cite{ADM}). Also in the
ADM case manifest covariance is lost because of the adoption of Lagrangian
and Hamiltonian variables which are not 4$-$tensors. In fact, ADM theory is
based on the introduction of a 3+1 decomposition of space-time which, by
construction, is foliation dependent, in the sense that it relies on a
peculiar choice of a family of GR frames for which "time" and "space"
transform separately, so that space-time is effectively split into the
direct product of a 1-dimensional time and a 3-dimensional space subsets
respectively \cite{alcu}. A quantum gravity theory constructed upon the ADM
Hamiltonian formulation of gravitational field leads to postulating a
quantum wave equation of Wheeler-DeWitt type \cite{dew}. The latter one is
expressed as an evolution Schr\"{o}dinger-like equation advancing the
dynamics of the wave function with respect to the coordinate-time $t$ of the
ADM foliation, which is not an invariant parameter. In addition, in the
absence of background space-time, the same equation carries a conceptual
problem related in principle to the definition of the same coordinate time,
which is simultaneously the dynamical parameter and a component of
space-time which must be quantized by solving the wave equation. This marks
a point of difference with respect to CQG theory and CQG-wave equation (see
Eq.(\ref{2}) below), which represents a dynamical evolution equation with
respect to an invariant (i.e., $4-$scalar) proper-time $s$ defined on the
prescribed background space-time, without introduction of any kind of
space-time foliation.

Another important approach is the one exemplified by the choice of so-called
Ashtekar variables, originally identified respectively with a suitable
self-dual spinorial connection (the generalized coordinates) and their
conjugate momenta (see Refs.\cite{Ashtekar1986, Ashtekar1987}). Ashtekar
variables provide an alternative canonical representation of General
Relativity, and this choice is at the basis of the so-called
\textquotedblleft loop representation of quantum general
relativity\textquotedblright\ \cite{Jacobson-Smolin1988} usually referred to
as "loop quantum gravity" (LQG) and first introduced by Rovelli and Smolin
in 1988-1990 \cite{Rovelli-Smolin1988,Rovelli-Smolin1990} (see also Ref.\cite%
{Rovelli1991}). Nevertheless, also the Ashtekar variables can be shown to be
by construction intrinsically manifestly non-tensorial in character. The
basic consequence is that also the canonical representation of Einstein
field equations based on these variables, as well as ultimately also LQG
itself, violates the principle of manifest covariance. In contrast, in the
framework of CQG-theory the choice of Hamiltonian state and quantum
variables satisfies manifest covariance, whereby the dynamical variables are
expressed by means of $4-$tensor quantities.

However, despite these considerations, it must be stressed that both the
canonical approach and CQG-theory can be regarded also complementary from a
certain point of view. This because they exhibit distinctive physical
properties associated with two canonical Hamiltonian structures underlying
General Relativity itself. The corresponding Hamiltonian flows, however, are
different, being referred to an appropriate coordinate-time of space-time
foliation in the canonical approach, and to a suitable invariant proper-time
in the present theory. As a consequence, the physical interpretation of
quantum theories of General Relativity build upon these Hamiltonian
structures remain distinctive. The CQG-theory in fact reveals the possible
existence of a discrete spectrum of metric tensors having non-vanishing
momenta at quantum level, while canonical approaches deal with the quantum
discretization of single space-time hypersurfaces implied by space-time
foliation.

Let us now consider the covariant approaches to quantum gravity \cite%
{Ashtekar1974,Weinber1972,DeWitt1972}. In this case the usual strategy is to
split the space-time metric tensor $g_{\mu \nu }$ in two parts according to
the decomposition of the type $g_{\mu \nu }=\eta _{\mu \nu }+h_{\mu \nu }$,
where $\eta _{\mu \nu }$ is the background metric tensor defining the
space-time geometry (usually identified with the flat background), and $%
h_{\mu \nu }$ is the dynamical field (deviation field) for which
quantization applies. From the conceptual point of view there are some
similarities between the literature covariant approaches and the
manifestly-covariant quantum gravity theory adopted here. The main points of
contact are: 1) the adoption of $4-$tensor variables, without invoking any
space-time foliation; 2) the implementation of a first-quantization
approach, in the sense that there exists by assumption a continuum classical
background space-time with a geometric connotation, over which the relevant
quantum fields are dynamically evolving; 3) the adoption of superabundant
variables, which in the two approaches are identified with the sets ($\eta
_{\mu \nu },h_{\mu \nu }$) and ($\widehat{g}_{\mu \nu },g_{\mu \nu }$)
respectively.

It is important nevertheless to emphasize the relevant differences existing
with respect to literature covariant approaches. First of all, CQG-theory is
intrinsically non-perturbative in character, so that the background metric
tensor can be identified with an arbitrary continuum solution of the
Einstein equations (not necessarily the flat space-time), while \textit{a
priori} the canonical variable $g_{\mu \nu }$ is not required to be
necessarily a perturbation field. On the other hand, a decomposition of the
type (\ref{TENSOR DECOMPOSITION}) resembling the one invoked in covariant
literature approaches can always be introduced \textit{a posteriori} for the
implementation of appropriate analytical solution methods, like GLP theory
proposed here or the analytical evaluation of discrete-spectrum quantum
solutions discussed in Ref.\cite{noi6}. Second, the present theory is
constructed starting from the DeDonder-Weyl manifestly-covariant approach
\cite{donder,weyl}. As a consequence CQG-theory is based on a variational
formulation which relies on the introduction of a synchronous variational
principle for the Einstein equations first reported in Ref.\cite{noi1}. This
represents a unique feature of manifestly-covariant quantum gravity theory,
since previous literature is actually based on the adoption of asynchronous
variational principles, i.e., in which the invariant volume element is
considered variational rather than prescribed. As shown in\ Ref.\cite{noi1}
it is precisely the synchronous principle which allows the distinction
between variational and extremal (or prescribed) metric tensors, and the
consequent introduction of non-vanishing canonical momenta. The same feature
has also made possible the formulation of manifestly-covariant classical
Lagrangian, Hamiltonian and Hamilton-Jacobi theories of General Relativity
and the corresponding subsequent manifestly-covariant quantum theory. Third,
in CQG-theory superabundant unconstrained variables are implemented, while
the same covariant quantization holds with respect to a $4-$dimensional
space-time, with no extra-dimensions being required for its prescription.

Finally, regarding covariant quantization, a further interesting comparison
concerns the Batalin-Vilkovisky formalism originally developed in Refs.\cite%
{BV2,BV3,BV4,BV5}. This method\ is usually implemented for the quantization
of gauge field theories and topological field theories in Lagrangian
formulation \cite{BV7,BV6,BV1}, while the corresponding Hamiltonian
formulation can be found in Ref.\cite{BFV1}. Further critical aspects of the
Batalin-Vilkovisky formalism can be found for example in Ref.\cite{BV8}. In
the case of the gravitational field it has been formerly applied of
perturbative quantum gravity to treat constraints arising from initial
metric decomposition (i.e., in reference with the so-called gauge-fixing and
ghost terms). Its basic features are the adoption of an asynchronous
Lagrangian variational principle of General Relativity \cite{noi1},\ the use
of superabundant canonical variables and the consequent introduction of
constraints. These features mark the main differences with CQG-theory, which
is non-perturbative, constraint-free and follows from the synchronous
Lagrangian variational principle defined in Ref.\cite{noi1}.

In view of these considerations, CQG-theory can be said to realize at the
same time both a canonical and a manifestly-covariant quantization method,
in this way establishing a connection with\ former canonical and covariant
approaches. Nevertheless, a number of conceptual new features of the present
theory depart in several ways from previous literature. This conclusion is
supported by the analytical results already established by CQG-theory and
presented in Refs.\cite{noi6,noi7}, which concern the existence of invariant
discrete-energy spectrum for the quantum gravitational field, the graviton
mass estimate associated with a non-vanishing cosmological constant and the
validity of Heisenberg inequalities.

Extending further these results, in the following a trajectory-based
representation of CQG-theory is developed, which permits the analytical
construction of generally non-stationary solutions of the CQG-wave equation.
Previous efforts to construct Bohmian representations of canonical quantum
gravity and applications to cosmology have been pursued in the past
literature. These are typically\textbf{\ }based on the Wheeler-DeWitt
quantum equation. Relevant progresses in this directions can be found for
example in Refs.\cite{x1,x2,x3,x4}, where conceptual features/differences
characterizing the Bohmian approach to quantum gravity (in terms of
trajectories) with respect to previous customary approaches were clearly
stated. The GLP representation of CQG-theory proposed in the present paper
shares the conceptual advantages of adopting a Bohmian approach to quantum
physics. On the other hand it differs from the mentioned literature in that
it is built upon CQG-theory, which is manifestly covariant contrary to the
Wheeler-DeWitt equation, and more important because it has a stochastic
character, namely in the sense that single Bohmian trajectories are replaced
by ensembles of stochastic trajectories with prescribed probability density.

\section{3 - Eulerian representation}

In this section the basic formalism of CQG-theory formulated in Refs.\cite%
{noi5,noi6} is recalled. Starting point of CQG-theory is the realization of
the quantum-wave function which, for an arbitrary prescribed background
space-time $\left( \mathbf{Q}^{4},\widehat{g}\right) $, determines the
CQG-quantum state. In analogy with non-relativistic quantum mechanics, this
can be first prescribed in the so-called Eulerian form. In this picture the
state is assumed to depend on two sets of independent variables,
respectively represented, first, by suitable configuration-space Lagrangian
variables and, second, by the space-time coordinates and time. In the
present case these are identified with the continuum field variables
(Lagrangian coordinates) $g\equiv \left\{ g_{\mu \nu }\right\} $ and,
respectively, by the 4-position $r\equiv \left\{ r^{\mu }\right\} $ and the
backgroundthe background space-time proper-time $s$, so that the wave
function takes generally the form $\psi \equiv \psi (g,r,s),$ where a
possible explicit dependence in terms of the background metric tensor $%
\widehat{g}$\ is understood. Regarding the notations, first$\ g=\left\{
g_{\mu \nu }\right\} $ spans the quantum configuration space $U_{g}$ of the
same wave-function, i.e., the set on which the associated quantum PDF $\rho
(g,r,s)=\left\vert \psi (g,r,s)\right\vert ^{2}$ is prescribed. Second, $%
g=\left\{ g_{\mu \nu }\right\} $\ is realized by means of real symmetric
tensors, so that $U_{g}$ is a $10-$dimensional real vector space, namely $%
U_{g}\subseteq
\mathbb{R}
^{10}$. Third, in the whole time-axis $I\equiv
\mathbb{R}
,$ $r\equiv \left\{ r^{\mu }\right\} $ denotes the instantaneous $4-$%
position of suitably-prescribed space-time trajectories $r=r(s)$,\textbf{\ }%
while the explicit $s-$dependence includes also the possible dependence (of $%
\psi $) in terms of the corresponding tangent $4-$vector, i.e., $t(s)\equiv
\left\{ t^{\mu }(s)\right\} \equiv \frac{dr^{\mu }(s)}{ds}$. Here $\frac{d}{%
ds}$\ identifies the total covariant $s-$derivative operator
\begin{equation}
\frac{d}{ds}\equiv \left. \frac{d}{ds}\right\vert _{r}+\left. \frac{d}{ds}%
\right\vert _{s},  \label{s-derivative}
\end{equation}%
with$\left. \frac{d}{ds}\right\vert _{r}\equiv \left. \frac{\partial }{%
\partial s}\right\vert _{r}$\ and $\left. \frac{d}{ds}\right\vert _{s}\equiv
t^{\alpha }\nabla _{\alpha }$\ being the covariant $s-$derivatives\
performed at constant $r\equiv \left\{ r^{\mu }\right\} $\ and constant $s$
respectively. A realization of the parametrization $\psi \equiv \psi (g,r,s)$
is provided by the geodetics of the metric field tensor $\widehat{g}\equiv
\widehat{g}(r)$, namely the integral curves of the initial-value problem
\cite{noi5,noi6}
\begin{equation}
\left\{
\begin{array}{c}
\frac{dr^{\mu }(s)}{ds}=t^{\mu }(s), \\
\frac{Dt^{\mu }(s)}{Ds}=0, \\
r^{\mu }(s_{o})=r_{o}^{\mu }, \\
t^{\mu }(s_{o})=t_{o}^{\mu },%
\end{array}%
\right.   \label{5}
\end{equation}%
with $\left( r_{o}^{\mu },t_{o}^{\mu }\right) $ denoting respectively
arbitrary initial $4-$position of $\left\{ \mathbf{Q}^{4},\widehat{g}%
\right\} $ and a corresponding (arbitrary) tangent $4-$vector, while the
standard connections in the covariant derivative $\frac{D}{Ds}$\ are
prescribed again in terms the background metric tensor $\widehat{g}(r)$.
Since each point $r^{\mu }\equiv r^{\mu }(s)$ can be crossed by infinite
arbitrary geodetics having different tangent $4-$vectors $t^{\mu }\equiv
t^{\mu }(s)$ it follows that the wave-function parametrization $\psi \equiv
\psi (g,r,s)$ may generally depend explicitly on the choice of the
geodetics, i.e., on $t^{\mu }$ too. In particular, $\psi (g,r,s)$ will be
assumed to contain the following smooth dependences:

1) \emph{Explicit }$g-$\emph{dependence: }$\psi (g,r,s)$ is assumed to be a $%
C^{(2)}-$smoothly differentiable complex function of the continuum
Lagrangian variables $g=\left\{ g_{\mu \nu }\right\} $.

2)\emph{\ Explicit and implicit }$s-$\emph{dependences: }$\psi (g,r,s)$ may
depend both explicitly and implicitly on $s$. The implicit dependence occurs
via $r(s)$ and $t(s)$ and therefore also in terms of the prescribed metric
tensor through its explicit spatial dependence $\widehat{g}(r)$. These $s-$%
dependences will all be assumed to realize in terms of $\psi (g,r,s)$ a $%
C^{(1)}-$smoothly differentiable function of $s$.

The next step is the identification of the quantum-wave equation which
determines the CQG state $\psi (g,r,s)$.\ This task is achieved by means of
the \emph{CQG-wave equation }\cite{noi6}. Written again in the Eulerian
form, the latter is realized by the initial-value problem%
\begin{equation}
\left\{
\begin{array}{c}
i\hslash \frac{\partial }{\partial s}\psi (g,r,s)=\left[ H_{R},\psi (g,r,s)%
\right] \equiv H_{R}\psi (g,r,s), \\
\psi (g,r(s_{o})=r_{o},s_{o})=\psi _{o}(g,r_{o}),%
\end{array}%
\right.   \label{2}
\end{equation}%
where in the first equation the squared-brackets denote the quantum
commutator in standard notation, while the operator $\frac{\partial }{%
\partial s}$ appearing in the scalar equation (\ref{2}) in the Eulerian
representation coincides with $\frac{\partial }{\partial s}\equiv \frac{d}{ds%
},$ being $\frac{d}{ds}$\ the total covariant $s-$derivative (\ref%
{s-derivative}) again prescribed in terms of the background metric tensor $%
\widehat{g}(r)$. Notice that the initial-value problem (\ref{2}) can be
represented equivalently in terms of the initial quantum fluid fields
\begin{equation}
\left\{
\begin{array}{c}
\rho (g,r(s_{o})=r_{o},s_{o})=\rho _{o}(g,r_{o}), \\
S^{(q)}(g,r(s_{o})=r_{o},s_{o})=S_{o}^{(q)}(g,r_{o}).%
\end{array}%
\right.   \label{initial fluid fields}
\end{equation}%
As such, provided $\psi (g,r,s)$\ is suitably smooth the solution of Eqs.(%
\ref{2}) is unique. Thus, Eqs.(\ref{2}) realize a hyperbolic evolution
equation, i.e., a first-order PDE with respect to the proper time $s$. In
the same equation\ $H_{R}$ denotes the quantum Hamiltonian operator
characteristic of CQG-theory, to be expressed in terms of the relevant
quantum momentum operator, namely $\pi _{\mu \nu }^{(q)}=-\frac{i\hbar }{%
\alpha L}\frac{\partial }{\partial g^{\mu \nu }}$. Here the partial
derivative is performed keeping constant all remaining variables appearing
in $\psi (g,r,s)$, while $L$ and $\alpha $ are respectively a
suitably-defined $4-$scalar scale-length and a dimensional $4-$scalar
parameter related to the universal constant $\kappa =\frac{c^{3}}{16\pi G}$
(see again Ref.\cite{noi6}). Then, the quantum Hamiltonian operator $H_{R}$
takes the form
\begin{equation}
H_{R}\equiv T_{R}^{(q)}+V(g,r,s),
\end{equation}%
with $T_{R}^{(q)}(x,\widehat{g})\equiv \frac{1}{2\alpha L}\pi _{\mu \nu
}^{(q)}\pi ^{(q)\mu \nu }$ and $V(g,r,s)$\textbf{\ }being respectively the
effective kinetic energy operator and the effective potential energy%
\begin{equation}
\left\{
\begin{array}{c}
V(g,r,s)\equiv \sigma V_{o}\left( g\right) +\sigma V_{F}\left( g,r,s\right) ,
\\
V_{o}\left( g\right) \equiv \alpha Lh\left[ g^{\mu \nu }\widehat{R}_{\mu \nu
}-2\Lambda \right] , \\
V_{F}\equiv \frac{\alpha L}{k}hL_{F}\left( g,r,s\right) ,%
\end{array}%
\right.   \label{FINAL}
\end{equation}%
with $V_{o}\left( g\right) $ and $V_{F}\left( g,r,s)\right) $ identifying
the vacuum and external effective contributions to the effective potential $%
V(g,r,s)$. Here the notation is given according to Ref.\cite{noi6}. Thus,
all hatted quantities are evaluated with respect to the background metric
tensor $\widehat{g}$\ only while the multiplicative $4-$scalar gauge
function $\sigma $\ is taken to be $\sigma =-1$. In addition, $V\equiv
V(g,r,s)$\ itself is determined up to an arbitrary additive gauge
transformation of the form $V\rightarrow V^{\prime }=V-\left. \frac{d}{ds}%
F(g,r(s)s)\right\vert _{g}$, being $F(g,r(s)s)$\ a $4-$scalar function of
the form%
\begin{equation}
F(g,r(s)s)=g_{\mu \nu }G^{\mu \nu }(\widehat{g},r(s),s)+F_{1}(\widehat{g}%
,r(s),s),  \label{prescription of the gauge functrion}
\end{equation}%
with $G^{\mu \nu }(\widehat{g},r(s),s)$\ and $F_{1}(\widehat{g},r(s),s)$
denoting respectively a $4-$tensor and a $4-$scalar smoothly differential
real gauge, i.e., arbitrary, functions. A characteristic element of
CQG-theory is the quantity\textbf{\ }$h\equiv h(g)$ first introduced in Ref.%
\cite{noi1}.\textbf{\ }The\ prescription of $h(g)$\ is obtained in terms of
a polynomial function of $g\equiv \widehat{g}+\delta g$,\ with $\delta g$\
being an in principle arbitrary variational displacement so that according
to the same reference (see also Ref. \cite{noi1}):%
\begin{equation}
h(g)=2-\frac{1}{4}\left( \widehat{g}^{\alpha \beta }+\delta g^{\alpha \beta
}\right) \left( \widehat{g}^{\mu \nu }+\delta g^{\mu \nu }\right) \widehat{g}%
_{\alpha \mu }\widehat{g}_{\beta \nu }.  \label{ORIGINAL PRESCR OF h}
\end{equation}

As a final remark, we notice that the Eulerian CQG-state defined by the
complex function $\psi (g,r,s)$ can always be cast in the form of an
exponential representation via the Madelung representation recalled above.\
Elementary algebra \cite{noi6,noi7} then shows that, based on the
quantum-wave equation (\ref{2}), the same quantum fluid fields necessarily
fulfill the corresponding set of Eulerian CQG-quantum hydrodynamic
equations. In the Eulerian representation, upon identifying again $\frac{d}{%
ds}$\ with the total covariant $s-$derivative operator (\ref{s-derivative}),
these are realized respectively by the continuity and quantum
Hamilton-Jacobi equations:%
\begin{equation}
\left\{
\begin{array}{c}
\frac{d\rho (g,r,s)}{ds}+\frac{\partial }{\partial g_{\mu \nu }}\left( \rho
(g,r,s)V_{\mu \nu }(g,r,s)\right) =0, \\
\frac{dS^{(q)}(g,r,s)}{ds}+H_{c}(g,r,s)=0,%
\end{array}%
\right.  \label{QHE-EULERIAN}
\end{equation}%
which represent a set of evolution PDEs for the quantum fluid fields $\rho
(g,r,s)$\ and $S^{(q)}(g,r,s)$.\ Notice that in the previous equations $%
V_{\mu \nu }(q,s)$ and $H_{c}(g,r,s)$ denote respectively the\emph{\ tensor
"velocity" field }$V_{\mu \nu }(g,r,s)=\frac{1}{\alpha L}\frac{\partial
S^{(q)}(g,r,s)}{\partial g^{\mu \nu }}$ and the\emph{\ effective quantum
Hamiltonian density}%
\begin{equation}
H_{c}(g,r,s)=\frac{1}{2\alpha L}\frac{\partial S^{(q)}(g,r,s)}{\partial
g^{\mu \nu }}\frac{\partial S^{(q)}(g,r,s)}{\partial g_{\mu \nu }}%
+V_{QM}(g,r,s)+V(g,r,s),  \label{prescription-2}
\end{equation}%
with%
\begin{equation}
T=\frac{1}{2\alpha L}\frac{\partial S^{(q)}(g,r,s)}{\partial g^{\mu \nu }}%
\frac{\partial S^{(q)}(g,r,s)}{\partial g_{\mu \nu }}
\label{effective kinetic energy}
\end{equation}%
being the \emph{effective kinetic energy}. In addition, $V(g,r,s)$\ and $%
V_{QM}(g,r,s)$ identify respectively the effective potential density (\ref%
{FINAL}) and\ the Bohm effective quantum potential%
\begin{equation}
V_{QM}(g,r,s)\equiv -\frac{\hslash ^{2}}{8\alpha L}\frac{\partial \ln \rho
(g,r,s)}{\partial g^{\mu \nu }}\frac{\partial \ln \rho (g,r,s)}{\partial
g_{\mu \nu }}-\frac{\hslash ^{2}}{4\alpha L}\frac{\partial ^{2}\ln \rho
(g,r,s)}{\partial g_{\mu \nu }\partial g^{\mu \nu }},  \label{BOHM-potential}
\end{equation}%
or equivalently $V_{QM}(g,r,s)\equiv \frac{\hslash ^{2}}{8\alpha L}\frac{%
\partial \ln \rho (g,r,s)}{\partial g^{\mu \nu }}\frac{\partial \ln \rho
(g,r,s)}{\partial g_{\mu \nu }}-\frac{\hslash ^{2}}{4\alpha L}\frac{\partial
^{2}\rho (g,r,s)}{\rho \partial g_{\mu \nu }\partial g^{\mu \nu }}.$

\section{4 - Lagrangian-path (Bohmian) representation}

It is well known that in the non-relativistic framework the Bohmian\
interpretation of quantum mechanics provides the corresponding
trajectory-based Lagrangian Path representation (\emph{LP-representation})
of the Schroedinger quantum-wave equation (see Ref.\cite{Holland,wyatt2005}
for a review of the topic). The intrinsic similarity with the CQG-wave
equation suggests that an analogous Lagrangian representation is possible
also for the same equation, so that - as a consequence - a "Bohmian"
trajectory-based interpretation can be achieved in the context of CQG-theory
too. In both cases, in fact, the Lagrangian representation is based on the
introduction of a suitable family of configuration-space trajectories, or
Lagrangian Paths (LP), which for each "point" of the appropriate quantum
configuration space are unique.

In the context of CQG-theory the LP-representation involves the introduction
for all $s\in I$\ of the correspondence (\ref{LP}), with $\delta g_{\mu \nu
}\equiv \delta g_{L\mu \nu }(s)\in U_{g}$\textbf{\ }belonging to a suitable
curve$\ \left\{ g_{L}(s),\forall s\in I\right\} $\ of the configuration
space $U_{g}$\ denoted as \emph{Lagrangian path}. As a consequence each LP
is identified with a well-defined characteristics associated with the tensor
velocity field $V_{\mu \nu }(g,r,s)$. For definiteness, based on the tensor
decomposition (\ref{TENSOR DECOMPOSITION}), the LP-representation involves%
\textbf{\ }parametrizing all quantum fields,\ and in particular the quantum
state, in terms of\ $g_{L\mu \nu }(s)$\ thus letting $\psi \equiv \psi
(g_{L\mu \nu }(s),r(s),s)$. As such, $\delta g_{L\mu \nu }(s)$ is
constructed in such a way that its "tangent"\ coincides with the local value
of the tensor velocity field $V_{\mu \nu }$, namely so that they fulfill the
initial-value problem%
\begin{equation}
\left\{
\begin{array}{c}
\frac{D}{Ds}g_{L\mu \nu }(s)=V_{\mu \nu }(g_{L}(s),s), \\
g_{\mu \nu }(s_{o})=g_{\mu \nu }^{(o)}.%
\end{array}%
\right.   \label{LP-1}
\end{equation}%
Here $\frac{D}{Ds}$ identifies the \emph{LP-derivative} (or \emph{covariant }%
$s-$\emph{derivative}) realized by the operator%
\begin{equation}
\frac{D}{Ds}\equiv \left. \frac{d}{ds}\right\vert _{\delta g_{L\mu \nu
}(s)}+V_{\mu \nu }(g_{L}(s),s)\frac{\partial }{\partial \delta g_{L\mu \nu }}%
,  \label{LP-operator}
\end{equation}%
where\ the two terms on the rhs\ of Eq.(\ref{LP-operator}) identify
respectively the\textbf{\ }\emph{covariant }$s-$\emph{derivative performed
at constant} $\delta g_{L\mu \nu }\equiv \delta g_{L\mu \nu }(s)$, namely%
\begin{equation}
\left. \frac{d}{ds}\right\vert _{\delta g_{L\mu \nu }(s)}\equiv \left. \frac{%
D}{Ds}\right\vert _{\delta g_{\mu \nu }}=\left[ \left. \frac{\partial }{%
\partial s}\right\vert _{r}+t^{\alpha }\nabla _{\alpha }\right] _{\delta
g_{L\mu \nu }},
\end{equation}%
and the convective derivative performed with respect to the Lagrangian
coordinates $\delta g_{L\mu \nu }(s)$\ while keeping constant\textbf{\ }$%
\widehat{g}(r)\equiv \left\{ \widehat{g}_{\mu \nu }(r(s))\right\} $. In view
of Eq.(\ref{TENSOR DECOMPOSITION}), Eq.(\ref{LP-1}) can be written as%
\begin{equation}
\text{ }\left\{
\begin{array}{c}
\frac{D}{Ds}\delta g_{L\mu \nu }(s)=V_{\mu \nu }(\widehat{g}(r)+\delta
g_{L}(s),s), \\
\delta g_{L\mu \nu }(s_{o})=\delta g_{\mu \nu }^{(o)}.%
\end{array}%
\right.   \label{LP-2}
\end{equation}%
As a consequence, Eq.(\ref{LP-2}) can be integrated to give%
\begin{equation}
\delta g_{L\mu \nu }(s)=\delta g_{\mu \nu
}^{(o)}+\int\limits_{s_{o}}^{s}ds^{\prime }V_{\mu \nu }(\widehat{g}%
(r)+\delta g_{L}(s^{\prime }),s^{\prime }),  \label{LP-3}
\end{equation}%
which determines the LP itself, namely the trajectory $\left\{
g_{L}(s),\forall s\in I\right\} \equiv \left\{ g_{L}(s)\equiv \widehat{g}%
(r)+\delta g_{L}(s),\forall s\in I\right\} $.\textbf{\ }However, if $H_{\mu
\nu }\equiv H_{\mu \nu }(\widehat{g}(r))$\ denotes an arbitrary
smoothly-differentiable tensor function of $\widehat{g}(r)$, it is obvious
that also the arbitrary additive tensor quantity of the form $\frac{D}{Ds}%
\left[ \delta g_{L\mu \nu }(s)+H_{\mu \nu }(\widehat{g}(r))\right] $
satisfies identically Eq.(\ref{LP-2}). Since uniqueness of the solution $%
\delta g_{L\mu \nu }(s)$\ given by Eq.(\ref{LP-3}) is warranted by
prescribing $\delta g_{\mu \nu }^{(o)},$\ the mapping\textbf{\ }%
\begin{equation}
g_{L\mu \nu }(s_{o})=g_{\mu \nu }^{(o)}\Leftrightarrow g_{L\mu \nu }(s)
\label{CDS-1}
\end{equation}%
identifies a classical dynamical system (CDS), i.e., a diffeomeorphism
mutually mapping in each other two arbitrary points $g_{L\mu \nu }(s_{o})$\
and $g_{L\mu \nu }(s)$\ which belong to the same LP. As a consequence, the
Liouville theorem warrants that the Jacobian determinant of the
transformation (\ref{CDS-1}) is%
\begin{equation}
\left\vert \frac{\partial \delta g_{L}(s)}{\partial \delta g_{L}(s_{o})}%
\right\vert =\exp \left\{ \int\limits_{s_{o}}^{s}ds^{\prime }\frac{\partial
V_{\mu \nu }(g_{L}(s^{\prime }),s^{\prime })}{\partial g_{L\mu \nu
}(s^{\prime })}\right\} .  \label{Jacobian of the CDS-1}
\end{equation}%
The Lagrangian representation of CQG-theory is then achieved by means of the
formal replacement $g\rightarrow g_{L}(s)$\ to be made in the quantum
wave-function, i.e., introducing in the CQG-wave equation (\ref{2}) the
\emph{LP-parametrization} $\psi =\psi (g_{L}(s),s)$ and similarly for the
quantum fluid fields, namely%
\begin{equation}
\left\{ \rho ,S^{(q)}\right\} \equiv \left\{ \rho
(g_{L}(s),s),S^{(q)}(g_{L}(s),s)\right\} .  \label{LP-fluid fields}
\end{equation}%
As a result, in terms of the tensor velocity field in the LP-representation,
namely $V_{\mu \nu }(g_{L}(s),s)\equiv \frac{1}{\alpha L}\frac{\partial
S^{(q)}(g_{L}(s),r(s),s)}{\partial \delta g_{L}^{\mu \nu }(s)}$, the quantum
hydrodynamic equations (\ref{QHE-EULERIAN})\ can be set at once in the
corresponding Lagrangian form.\textbf{\ }To obtain them one notices
preliminarily that%
\begin{equation}
\frac{D}{Ds}S^{(q)}(g_{L}(s),s)\equiv \frac{d}{ds}S^{(q)}(g_{L}(s),s)+V_{\mu
\nu }(g_{L}(s),s)\frac{\partial S^{(q)}(g_{L}(s),s)}{\partial g_{L\mu \nu
}(s)},
\end{equation}%
with $\frac{D}{Ds}$ and $\frac{d}{ds}$ identifying respectively\ the
LP-derivative (\ref{LP-operator}) and the total covariant $s-$derivative
operator (\ref{s-derivative}). As a consequence, the LP-representation of
the quantum fluid equations (\ref{QHE-EULERIAN}) is given respectively by
the PDEs%
\begin{equation}
\left\{
\begin{array}{c}
\frac{D}{Ds}\rho (g_{L}(s),s)=-\rho (g_{L}(s),s)\frac{\partial V_{\mu \nu
}(g_{L}(s),s)}{\partial g_{L\mu \nu }(s)}, \\
\frac{D}{Ds}S^{(q)}(g_{L}(s),s)=V_{\mu \nu }(g_{L}(s),s)\frac{\partial
S^{(q)}(g_{L}(s),s)}{\partial g_{L\mu \nu }(s)}-H_{c}(g_{L}(s),s),%
\end{array}%
\right.   \label{QHE-LAGRANGIAN}
\end{equation}%
where $H_{c}(g_{L}(s),s)$ identifies the effective quantum Hamiltonian
density (\ref{prescription-2}) parametrized in terms of $g_{L}(s)$. Thus, in
particular, the continuity equation (first equation in (\ref{QHE-LAGRANGIAN}%
)) can be formally integrated to give the\ LP-parametrized integral
continuity equation%
\begin{equation}
\rho (g_{L}(s),s)=\rho (g_{L}(s_{o}),s_{o})\exp \left\{
-\int\limits_{s_{o}}^{s}ds^{\prime }\frac{\partial V_{\mu \nu
}(g_{L}(s^{\prime }),s^{\prime })}{\partial g_{L\mu \nu }(s^{\prime })}%
\right\} ,  \label{INTEGRAL CONTINUITY eq.}
\end{equation}%
with $\rho (g_{L}(s_{o}),s_{o})\equiv \rho
(g_{L}(s_{o}),r(s_{o})=r_{o},s_{o})$ denoting the initial quantum PDF, namely%
\begin{equation}
\rho (g_{L}(s_{o}),r(s_{o})=r_{o},s_{o})=\rho _{o}(g_{L}(s_{o}),r_{o}).
\label{INITIAL PDF LP}
\end{equation}%
Together with Liouville theorem (\ref{Jacobian of the CDS-1}) this implies
therefore the conservation laws%
\begin{equation}
d(g_{L}(s))\rho (g_{L}(s),s)=d(g_{L}(s_{o}))\rho (g_{L}(s_{o}),s_{o}),
\label{CONS-LAW-1}
\end{equation}%
\begin{equation}
\int\limits_{U_{g}}d(g_{L}(s))\rho
(g_{L}(s),s)=\int\limits_{U_{g}}d(g_{L}(s_{o}))\rho (g_{L}(s_{o}),s_{o})=1,
\label{CONS-LAW-2}
\end{equation}%
which warrant,\ consistent with the quantum unitarity principle, the
conservation of the quantum probability in $U_{g}$.

We conclude this section noting that from a mathematical viewpoint the
Lagrangian formulation of CQG-theory is actually realized solely by the
LP-parametrized quantum hydrodynamic equations (\ref{QHE-LAGRANGIAN}).
Therefore the Lagrangian and Eulerian quantum hydrodynamic equations are
manifestly equivalent. This suggests that a Bohmian interpretation of the
Lagrangian-path representation of the CQG-theory is in principle possible.
However, just as in the case of the Schroedinger equation (see related
discussion in Ref.\cite{Tessarotto2016}), a basic difficulty of such an
interpretations lies in the uniqueness feature, and consequently the
intrinsic deterministic character, of each LP. Such a property, in fact,
appears potentially in contradiction with the notion of quantum measurement
holding in the context of CQG-theory and the validity of Heisenberg
inequalities \cite{noi7}.

\section{5 - Generalized Lagrangian-path representation}

The considerations indicated above lead us to introduce the notion of
Generalized Lagrangian Path (GLP) and of the corresponding
GLP-representation obtained in this way for the quantum wave-function and
quantum fluids fields. As anticipated above (see Introduction) this is
achieved by means of the introduction of a suitable set of intrinsically
non-unique and stochastic trajectories, to be referred to as generalized
Lagrangian paths (GLPs), in terms of which the quantum wave-equation, as
well as the corresponding set of quantum fluid fields and quantum
hydrodynamic equations, can be parametrized. In the context of CQG-theory
the mathematical problem of formulating its GLP-representation\textbf{\ }%
involves the introduction for all $s\in I$ of a suitable correspondence of
the type%
\begin{equation}
s\rightarrow \delta G_{L}(s),  \label{GLP-map}
\end{equation}%
referred to as \emph{GLP-map.} Then, upon invoking the tensor decomposition (%
\ref{TENSOR DECOMPOSITION 2}), a GLP is the curve $\left\{ G_{L}(s),\forall
s\in I\right\} $\ of the quantum configuration space $U_{g}$\ which is
defined by Eq.(\ref{GLP definition}) and is realized by the ensemble of
"points" of $U_{g}$\ spanned by the tensor field $G_{L}(s)\equiv G(s)$\ and
obtained varying $s\in I$.\textbf{\ }The underlying basic idea is therefore
to replace a single LP, prescribed in terms of a solution of the
initial-value problem (\ref{LP-1}), with an infinite set of stochastic
trajectories, each one identified with a single GLP and characterized by a
unique choice of a suitable stochastic tensor $\Delta g=\left\{ \Delta
g_{\mu \nu }\right\} $. This effectively involves introducing a
parameter-dependent mapping of the type%
\begin{equation}
\left\{ g_{L}(s),\forall s\in I\right\} \rightarrow \left\{ G_{L}(s),\forall
s\in I\right\} ,  \label{MAP to GLP}
\end{equation}%
whose realization depends on the prescription of $\Delta g=\left\{ \Delta
g_{\mu \nu }\right\} $. Then the GLP-map (\ref{MAP to GLP}) is realized by
means of the following two requirements.

\begin{itemize}
\item \emph{GLP Requirement \#1} \emph{-} The first one is realized by
prescribing $\delta G_{L\mu \nu }(s)$ in terms of the displacement tensor $%
\delta g_{L\mu \nu }(s)$ which is determined according to Eq.(\ref{TENSOR
DECOMPOSITION}). This yields\ therefore the identity%
\begin{equation}
G_{L\mu \nu }(s)=\widehat{g}_{\mu \nu }(r)+\delta g_{L\mu \nu }(s)-\Delta
g_{\mu \nu },
\end{equation}%
with $\Delta g$ denoting the \emph{stochastic displacement }$4-$\emph{tensor}%
\begin{equation}
\Delta g=g-G_{L}(s)\equiv \delta g-\delta G_{L}(s).
\label{DSIPALCEMENT 4-TENSOR}
\end{equation}%
Notice that here $g_{\mu \nu }=g_{L\mu \nu }(s)$, and hence $\delta g_{\mu
\nu }\equiv \delta g_{L\mu \nu }(s)$. As a consequence, it is understood
that $\Delta g$ must be endowed with a suitable stochastic PDF to be
suitably prescribed. In this regards, taking $\Delta g$ as an independent
stochastic variable, it is natural to assume that the same PDF should be a
stationary and spatially uniform probability distribution, i.e., a function
independent of $r,$ $s$\ as well as $\delta g_{L}(s)$,\ but still allowed to
depend in principle on the prescribed metric tensor $\widehat{g}_{\mu \nu
}(r)$.\textbf{\ }More precisely, this means assuming the same PDF to be
realized in terms of a smoothly differentiable and strictly positive
function of the form%
\begin{equation}
f=f\left( \Delta g,\widehat{g}\right) .  \label{STOCHASTIC PDF}
\end{equation}%
Hence the corresponding notion of stochastic average for an arbitrary smooth
function $X(\Delta g,r,s)$\ is prescribed in terms of the weighted integral%
\begin{equation}
\left\langle X(\Delta g,r,s)\right\rangle _{stoch}\equiv
\int\limits_{U_{g}}d(\Delta g)X(\Delta g,r,s)f\left( \Delta g,\widehat{g}%
\right) ,
\end{equation}%
to be performed on the configuration space $U_{g}$. In particular, besides
the prescription (\ref{STOCHASTIC PDF}), $f\left( \Delta g,\widehat{g}%
\right) $ should be prescribed so that the following stochastic averages are
also fulfilled:%
\begin{equation}
\left\{
\begin{array}{c}
\left\langle 1\right\rangle _{stoch}\equiv \int\limits_{U_{g}}d(\Delta
g)f\left( \Delta g,\widehat{g}\right) =1, \\
\left\langle \Delta g_{\mu \nu }\right\rangle _{stoch}\equiv
\int\limits_{U_{g}}d(\Delta g)\Delta g_{\mu \nu }f\left( \Delta g,\widehat{g}%
\right) )=\pm \widehat{g}_{\mu \nu }(r), \\
\sigma _{\Delta g}^{2}\equiv \left\langle \left( \Delta g-\left\langle
\Delta g\right\rangle _{stoch}\right) ^{2}\right\rangle _{stoch}\equiv  \\
\int\limits_{U_{g}}d(\Delta g)\left( \Delta g-\left\langle \Delta
g\right\rangle _{stoch}\right) ^{2}f\left( \Delta g,\widehat{g}\right)
=r_{th}^{2},%
\end{array}%
\right.   \label{CONSTRAINTS ON f(Dg)}
\end{equation}%
with $\left( \Delta g-\left\langle \Delta g\right\rangle _{stoch}\right)
^{2}\equiv \left[ \Delta g_{\mu \nu }-\left\langle \Delta g_{\mu \nu
}\right\rangle _{stoch}\right] \left[ \Delta g^{\mu \nu }-\left\langle
\Delta g^{\mu \nu }\right\rangle _{stoch}\right] $ and $\sigma _{\Delta g}$
denoting the standard deviation of $\Delta g$ to be identified with the
dimensionless $4-$scalar parameter $r_{th}^{2}>0$.\ Notice in addition that
here, for consistency with the same assumption (\ref{STOCHASTIC PDF}), $%
r_{th}^{2}$\ must be assumed to be a non-vanishing constant, i.e.,
independent of both $(r,s)$.

\item \emph{GLP Requirement \#2 - }The second one is obtained requiring that
$\Delta g=\left\{ \Delta g_{\mu \nu }\right\} $ is constant for all $s\in I$
and for an arbitrary Lagrangian Path, i.e., it is prescribed so that
identically for all $s,s_{o}\in I$ it occurs that%
\begin{equation}
\Delta g_{\mu \nu }(s)=\Delta g_{\mu \nu }(s_{o}).  \label{CONSTANT-DELTA-0}
\end{equation}%
Notice that here $\frac{D}{Ds}$ $\delta g_{L\mu \nu }(s)=\frac{D}{Ds}\delta
G_{L\mu \nu }(s)\equiv V_{\mu \nu }(G_{L}(s),\Delta g,s)),$ with $V_{\mu \nu
}(G_{L}(s),\Delta g,s)$\ being the\ tensor velocity field in the
GLP-representation,\ namely%
\begin{equation}
V_{\mu \nu }(G_{L}(s),\Delta g,s)=\frac{1}{\alpha L}\frac{\partial
S^{(q)}(G_{L}(s),\Delta g,s)}{\partial \delta g_{L}^{\mu \nu }(s)},
\label{GLP-parametrization of tensor velocity}
\end{equation}%
while $\frac{D}{Ds}$\ is the Lagrangian derivative defined above (see Eq.(%
\ref{LP-operator})).\textbf{\ }As a result, the constraint condition (\ref%
{CONSTANT-DELTA-0}) necessarily implies also that%
\begin{equation}
\frac{D}{Ds}\Delta g_{\mu \nu }\equiv \frac{D}{Ds}\delta g_{L\mu \nu }(s)-%
\frac{D}{Ds}\delta G_{L\mu \nu }(s)\equiv 0.  \label{CONSTANT-DELTA-1}
\end{equation}
\end{itemize}

As a consequences of Requirements \#1 and \#2, for all $s\in I$ the
correspondence in Eq.(\ref{GLP-map}) is uniquely established, in the sense
that, for each determination of the stochastic displacement $\Delta g,$ $%
G_{L}(s,\Delta g)\equiv G_{L}(s)$ belongs to a uniquely-prescribed curve$\
\left\{ G_{L}(s),\forall s\in I\right\} ,$ identifying a GLP\ which spans
the quantum configuration space $U_{g}$. More precisely, a generic GLP $%
\left\{ G_{L}(s),\forall s\in I\right\} $ is identified with the integral
curve determined by the GLP-initial-value problem%
\begin{equation}
\left\{
\begin{array}{c}
\frac{D}{Ds}\delta G_{L\mu \nu }(s)=V_{\mu \nu }(G_{L}(s),\Delta g,s), \\
\delta G_{L\mu \nu }(s_{o})=\delta g_{\mu \nu }^{(o)}-\Delta g_{\mu \nu }.%
\end{array}%
\right.   \label{GLP-initial-value problem}
\end{equation}%
In addition, here the map $G_{L}(s_{o})\Leftrightarrow G_{L}(s)$ defines
again a classical dynamical system with Jacobian determinant%
\begin{equation}
\left\vert \frac{\partial G_{L}(s)}{\partial G_{L}(s_{o})}\right\vert =\exp
\left\{ \int\limits_{s_{o}}^{s}ds^{\prime }\frac{\partial V_{\mu \nu
}(G_{L}(s^{\prime })+\Delta g,s^{\prime })}{\partial g_{L\mu \nu }(s^{\prime
})}\right\} .
\end{equation}

The ensemble of integral curves $\left\{ G_{L}(s),\forall s\in I\right\} $
obtained by varying $\Delta g$ in $U_{g}$ identifies therefore an infinite
set of GLP which are associated with the tensor velocity field $V_{\mu \nu
}(G_{L}(s)+\Delta g,s)$. One notices, however, that by construction
\begin{equation}
V_{\mu \nu }(G_{L}(s)+\Delta g,s)=V_{\mu \nu }(g_{L}(s),s).
\end{equation}%
As a consequence the same infinite set of GLP is actually associated \emph{%
with the}\textbf{\ }\emph{same local value of the tensor velocity field}%
\textbf{\ }$V_{\mu \nu }(g_{L}(s),s)$.\textbf{\ }Thus, in contrast with the
LP defined above (in terms of Eq.(\ref{LP-1})), this means that the GLP
which are associated with the local tensor velocity field $V_{\mu \nu
}(g_{L}(s),s)$ are non-unique (and actually infinite), each one being
determined by $\Delta g$. Precisely because the same trajectories are
stochastic and hence non-unique, such a feature is in principle compatible
with the possible interpretation of the GLPs as \emph{physical quantum
trajectories in the configurations space} $U_{g}$. Nevertheless, the
prerequisite for making actually possible such an interpretation is,
ultimately, the ontological equivalence of the GLP-parametrization for the
quantum state $\psi $\ with the "standard" Eulerian representation of the
same quantum wave-function.\textbf{\ }In other words, the adoption of the
GLP and in particular the prescription of the stochastic PDF $f(\Delta g)$\
associated with the same constant stochastic displacement tensor $\Delta g$\
(see Eq.(\ref{STOCHASTIC PDF})), should be possible leaving unchanged the
axioms of CQG-theory.

For definiteness, let us now pose the problem of introducing explicitly the
parametrization of the quantum fluid fields and the related
GLP-representation of the QHE. In principle, this can simply be obtained
from the corresponding LP-parametrization indicated above noting that $%
\delta g(s)=\Delta g+\delta G(s)$. However, in formal analogy with the
GLP-approach to non-relativistic quantum mechanics earlier indicated, a more
general parametrization in terms of the stochastic displacement tensor field
$\Delta g$, to be referred in the sequel as GLP-parametrization, is possible.%
\textbf{\ }This involves assuming that the CQG-wave function may be of the
type%
\begin{equation}
\psi =\psi (G_{L}(s),\Delta g,s),  \label{GLP-prescription-1}
\end{equation}%
i.e., to include also an explicit dependence in terms of $\Delta g\equiv
\left\{ \Delta g_{\mu \nu }\right\} $. Therefore, the corresponding
GLP-parametrization of the quantum fluid fields is taken of the form%
\begin{equation}
\left\{ \rho ,S^{(q)}\right\} _{(s)}\equiv \left\{ \rho (G_{L}(s),\Delta
g,s),S^{(q)}(G_{L}(s),\Delta g,s)\right\} .  \label{GLP-prescription-2}
\end{equation}%
Nevertheless, the quantum hydrodynamic equations (\ref{QHE-EULERIAN}) when
expressed in the GLP-parametrization remain formally analogous to those
obtained in the LP-parametrization (see Eqs.(\ref{QHE-LAGRANGIAN})), so that
the same equations must determine the map%
\begin{equation}
\left\{ \rho ,S^{(q)}\right\} _{(s_{o})}\equiv \left\{ \rho
_{o},S_{o}^{(q)}\right\} \rightarrow \left\{ \rho ,S^{(q)}\right\} _{(s)},
\label{GLP-MAP}
\end{equation}%
with $\left\{ \rho _{o},S_{o}^{(q)}\right\} $\ being suitable initial
quantum fluid fields. Hence, for consistency these should be again assumed
of the form%
\begin{equation}
\left\{ \rho _{o},S_{o}^{(q)}\right\} \equiv \left\{ \rho
_{o}(G_{L}(s_{o}),\Delta g),S_{o}^{(q)}(G_{L}(s_{o}),\Delta g)\right\} .
\label{INITIAL GLP-FLUID FIELDS}
\end{equation}%
In detail, in the GLP-representation the quantum hydrodynamic equations (\ref%
{QHE-LAGRANGIAN}) are\ now realized by the PDEs%
\begin{equation}
\left\{
\begin{array}{c}
\frac{D}{Ds}\rho (G_{L}(s),\Delta g,s)=-\rho (G_{L}(s),\Delta g,s)\frac{%
\partial V_{\mu \nu }(G_{L}(s),\Delta g,s)}{\partial g_{L\mu \nu }(s)}, \\
\frac{D}{Ds}S^{(q)}(G_{L}(s),\Delta g,s)=K_{c}(G_{L}(s),\Delta g,s),%
\end{array}%
\right.  \label{GLP-equations}
\end{equation}%
representing respectively the\ \emph{GLP-parametrized quantum continuity }%
and \emph{H-J equations}, where%
\begin{equation}
K_{c}(G_{L}(s),\Delta g,s)=V_{\mu \nu }(G_{L}(s),\Delta g,s)\frac{\partial
S^{(q)}(G_{L}(s,\Delta g),\Delta g,s)}{\partial g_{L\mu \nu }(s)}%
-H_{c}(G_{L}(s),\Delta g,s)
\end{equation}%
and $H_{c}(G_{L}(s),\Delta g,s)$ identifies now the effective quantum
Hamiltonian density (\ref{prescription-2}) expressed in terms of the
GLP-parametrization. Thus from Eq.(\ref{prescription-2}) it follows that%
\begin{equation}
H_{c}(G_{L}(s),\Delta g,s)=T(G_{L}(s),\Delta g,s)-V(G_{L}(s),\Delta
g,s)-V_{QM}(G_{L}(s),\Delta g,s),  \label{GLP-K_c}
\end{equation}%
with $T\equiv T(G_{L}(s),\Delta g,s),$ $V\equiv V(G_{L}(s),\Delta g,s)$\ and
$V_{QM}\equiv V_{QM}(G_{L}(s),\Delta g,s)$\textbf{\ }denoting now in terms
of the GLP-parametrization respectively the effective kinetic energy and
classical potential density given by Eqs.(\ref{effective kinetic energy}), (%
\ref{FINAL}) and\ the Bohm effective quantum potential (\ref{BOHM-potential}%
). Thus, regarding the representation of the effective potential energy $V,$
and in particular\ its vacuum contribution $V_{o}\equiv
V_{o}(G_{L}(s),\Delta g,s)$\ (see Eqs.(\ref{FINAL})),\textbf{\ }to be used
in the context of the GLP-approach, one notices that the displacement $4-$%
tensor $\delta g$\ entering the expression of the variational parameter (\ref%
{ORIGINAL PRESCR OF h}) remains non-unique. One notices that, due to its
arbitrariness, the displacement $4-$tensor\ can always be identified with $%
\delta g\equiv \Delta g$, being $\Delta g$\ the stochastic constant
displacement field tensor introduced above (see Eq.(\ref{SETTING})), so that
actually $h(g)$\ can be conveniently represented as\textbf{\ }%
\begin{equation}
h(\widehat{g}+\Delta g)=2-\frac{1}{4}\left( \widehat{g}^{\alpha \beta
}+\Delta g^{\alpha \beta }\right) \left( \widehat{g}^{\mu \nu }+\Delta
g^{\mu \nu }\right) \widehat{g}_{\alpha \mu }\widehat{g}_{\beta \nu },
\label{prescription of h}
\end{equation}%
while the vacuum effective potential becomes:%
\begin{equation}
V_{o}(G_{L}(s),\Delta g,s)\equiv \sigma \alpha Lh(\widehat{g}+\Delta g)\left[
\left( \widehat{g}_{pq}(s)+\Delta g_{pq}\right) \widehat{g}^{pq}(r)-2\right]
\Lambda .  \label{VACUUM EFFECTIVE POTENTIAL}
\end{equation}

Useful implications of the GLP-representation (\ref{GLP-prescription-1})-(%
\ref{GLP-prescription-2}) follow by inspection of the GLP-quantum continuity
equation (see first equation in (\ref{GLP-equations})) obtained above. The
first one follows by noting that the same equation implies also%
\begin{equation}
\frac{D}{Ds}\ln \rho (G_{L}(s),\Delta g,s)=-\frac{\partial V_{\mu \nu
}(G_{L}(s),\Delta g,s)}{\partial g_{L\mu \nu }(s)},
\end{equation}%
so that its formal integration generates the map $\rho (G_{L}(s_{o}),\Delta
g,s_{o})\rightarrow \rho (G_{L}(s),\Delta g,s),$ with $\rho (G_{L}(s),\Delta
g,s)$ denoting the \emph{proper-time evolved quantum PDF,}\textbf{\ }namely%
\begin{equation}
\rho (G_{L}(s),\Delta g,s)=\rho (G_{L}(s_{o}),\Delta g,s_{o})\exp \left\{
-\int\limits_{s_{o}}^{s}ds^{\prime }\frac{\partial V_{\mu \nu
}(G_{L}(s^{\prime }),\Delta g,s^{\prime })}{\partial g_{L\mu \nu }(s^{\prime
})}\right\} .  \label{SOLUTION OF CONTINUITY EQ}
\end{equation}%
Notice that the integration on the rhs is performed along the GLP-trajectory
$\left\{ G_{L}(s,\Delta g),\forall s\in I\right\} $,\ i.e., for a prescribed
constant stochastic displacement $4-$tensor $\Delta g,$\ while $\rho
(G_{L}(s_{o}),\Delta g,s_{o})$\ identifies the initial, and in principle
still arbitrary, PDF.\textbf{\ }The second implication concerns the quantum
H-J equation itself. In fact the formal solution (\ref{SOLUTION OF
CONTINUITY EQ}) permits to cast it in terms of an (implicit) equation for
the GLP-parametrized quantum phase-function $S^{(q)}(G_{L}(s),\Delta g,s)$
only.\textbf{\ }As a consequence, provided an explicit realization is
reached for the GLP-trajectory $\left\{ G_{L}(s),\forall s\in I\right\} $,
by solving the initial-value problem (\ref{GLP-initial-value problem}), the
same H-J equation should uniquely determine the corresponding solution $%
S^{(q)}(G_{L}(s),\Delta g,s)$\ as a real function of $\Delta g$\ and $s$\
only.\textbf{\ }A notable\ feature worth to be stressed here is about the
prescription of the same initial PDF $\rho (G_{L}(s_{o}),\Delta g,s_{o})$.
This manifestly generally differs from the one considered above in the case
of the LP-parametrization (see Eq.(\ref{INITIAL PDF LP})), where no explicit
$\Delta g-$dependences was assumed.\ In fact, consistent with the
GLP-parametrization introduced above (see Eq.(\ref{GLP-prescription-2})),
this is now taken of the form (\ref{INITIAL GLP-FLUID FIELDS}). This means
that it may include in particular an admissible choice for the initial PDF
provided by a probability density of the form%
\begin{equation}
\rho (G_{L}(s_{o}),\Delta g,s_{o})=\rho _{o}(\Delta g+\widehat{g}(r_{o})),
\label{initial quantum pdf}
\end{equation}%
with $\rho (\Delta g+\widehat{g}(r_{o}))$\ to be determined as indicated
below.

\section{6 - GLP approach: determination of the stochastic PDF for $\Delta g$
and of the quantum PDF}

The problem addressed in this section is twofold. First, it concerns the
identification of the stochastic probability density $f(\Delta g,\widehat{g}%
_{\mu \nu })$ which is associated with the stochastic displacement tensor
field $\Delta g\equiv \left\{ \Delta g_{\mu \nu }\right\} $ and is\
consistent with the requirements indicated above, i.e., Eq.(\ref{STOCHASTIC
PDF}), together with the aforementioned constraint conditions (\ref%
{CONSTRAINTS ON f(Dg)}). Second, it deals with the prescription of the
CQG-probability density, in particular the initial one $\rho _{o}$, to be
adopted in the GLP-parametrization, see Eq.(\ref{initial fluid fields})\ as
well Eqs.(\ref{GLP-prescription-2}) and (\ref{initial quantum pdf}) above.
In fact, both prescriptions should be actually regarded as mandatory
prerequisites for the consistency of the GLP-representation and its
ontological equivalence with the corresponding Eulerian representation of
CQG-theory. In this Section we intend to show that the two issues are
actually intrinsically related.

In particular we aim to prove that the initial quantum PDF can be prescribed
in such a way that it coincides with a shifted Gaussian PDF, such a choice
being consistent with the principle of entropy maximization (PEM), i.e.,%
\textbf{\ }determined so to maximize the initial Boltzmann-Shannon entropy
associated with the initial PDF. As a consequence, the same initial PDF is
shown to satisfy suitable symmetry properties (see Proposition \#1).
Furthermore the problem is posed of the determination of the quantum
expectation values evaluated with respect to the GLP-parametrized quantum
PDF.\textbf{\ }As a result, for arbitrary observables which are identified
with ordinary tensor functions, equivalent representations of the
GLP-quantum expectation values\ are pointed out (Proposition\ \#2).\textbf{\
}A notable related implication refers to the physical interpretation of
CQG-theory arising in such a context which is analogous to the so-called
emergent gravity picture of quantum gravity. This follows by noting that by
suitable prescription of the initial quantum PDF the background metric
tensor $\widehat{g}(r(s))$ is\ uniquely determined, at any arbitrary
proper-time $s,$\ in terms of an appropriate expectation value of the
quantum PDF (see Proposition \#3).

\subsection{Prescription of the stochastic PDF}

The two topics indicated above actually have a unique solution. This follows
at once provided the axiomatic setting of CQG-theory is invoked.\ Let us
consider, in fact, the problem of the determination of $f(\Delta g,\widehat{g%
})$. In the context of CQG-theory, as in the case of Quantum Mechanics (see
related discussion in Ref.\cite{Tessarotto2016}), the independent
prescription of $f(\Delta g,\widehat{g})$ potentially may amount to the
introduction of an additional axiom, thus possibly giving rise to additional
conceptual difficulties related to the notions of quantum measurement and
quantum expectation values. In order to overcome the issue, while leaving
unaffected the axioms of CQG-theory earlier introduced in Ref.\cite{noi6}
and, at the same time, warranting the ontological equivalence indicated
above, the only possible choice for $f(\Delta g,\widehat{g})$ is that it
coincides with the initial quantum PDF $\rho _{o}$. This means also,\ of
course, that $\rho _{o}$ must be necessarily of the type (\ref{initial
quantum pdf}), namely such that%
\begin{equation}
f(\Delta g,\widehat{g})\equiv \rho _{o}(\Delta g\pm \widehat{g}(r_{o})),
\label{PRESCRIPTION OF f(Dg)}
\end{equation}%
and therefore fulfilling also the constraint conditions indicated above (see
Eqs.(\ref{CONSTRAINTS ON f(Dg)})). Incidentally, as explained below, from
the conceptual viewpoint this choice exhibits remarkable features.

\subsection{The initial quantum\ PDF $\protect\rho _{o}$ and its invariance
property}

The first one, as a specific application of the GLP formalism, concerns the
prescription itself of the initial quantum PDF.\textbf{\ }In fact, in
validity of the identification (\ref{PRESCRIPTION OF f(Dg)}), the
constraints (\ref{CONSTRAINTS ON f(Dg)}) included in the \emph{Requirement
\#1} indicated above actually uniquely prescribe the form of the initial PDF
$\rho _{o}(\Delta g\pm \widehat{g}(r_{o}))$. In fact, let us introduce for
definiteness the Boltzmann-Shannon entropy associated with the same PDF,
which is provided by the functional
\begin{equation}
S(\rho _{o}(\Delta g+\widehat{g}(r_{o})))=-\int_{U_{g}}d(\Delta g)\rho
_{o}(\Delta g+\widehat{g}(r_{o}))\ln \rho _{o}(\Delta g+\widehat{g}(r_{o})),
\label{B-S entropy}
\end{equation}%
with $\rho _{o}(\Delta g,\widehat{g}_{\mu \nu }(r))\equiv f(\Delta g+%
\widehat{g}_{\mu \nu }(r))$ being assumed to satisfy the same constraint
equations indicated above (i.e., Eqs.(\ref{CONSTRAINTS ON f(Dg)}). Then, one
can show that the PDF $\rho _{o}(\Delta g+\widehat{g}(r_{o}))$\ which
fulfills the so-called Principle of Entropy Maximization (PEM, Jaynes 1957),
namely maximizes $S(\rho _{o}(\Delta g+\widehat{g}(r_{o})))$\ when subject
to the same constraints, is unique. Straightforward algebra shows that in
the whole configuration domain $U_{g}$\ it coincides with the PDF
\begin{equation}
\rho _{o}(\Delta g\pm \widehat{g}(r_{o}))=\frac{1}{\pi ^{5}r_{th}^{10}}\exp
\left\{ -\frac{\left( \Delta g\pm \widehat{g}(r_{o})\right) ^{2}}{r_{th}^{2}}%
\right\} \equiv \rho _{G}(\Delta g\pm \widehat{g}(r_{o})),
\label{REALIZATION OF THE INITIAL ODF}
\end{equation}%
with $\rho _{G}(\Delta g\pm \widehat{g}(r_{o}))$ denoting a \emph{shifted
Gaussian PDF} in which both $r_{th}^{2}$\ and $\left( \Delta g\pm \widehat{g}%
(r_{o})\right) ^{2}$\ are $4-$scalars, and in particular $r_{th}^{2}$\ is a
constant independent of $(r,s)$, while%
\begin{equation}
\left( \Delta g\pm \widehat{g}(r_{o})\right) ^{2}\equiv \left( \Delta g\pm
\widehat{g}(r_{o})\right) _{\mu \nu }\left( \Delta g\pm \widehat{g}%
(r_{o})\right) ^{\mu \nu }.  \label{INVARIANCE PROPERTY}
\end{equation}

Therefore, we conclude that the Gaussian PDF (\ref{REALIZATION OF THE
INITIAL ODF}) realizes \emph{the most likely PDF}, i.e., the one which -
when subject to the constraints (\ref{CONSTRAINTS ON f(Dg)}) - maximizes the
Boltzmann-Shannon entropy $S(\rho _{o}(\Delta g+\widehat{g}(r_{o})))$ in Eq.(%
\ref{B-S entropy}).

Let us now denote with%
\begin{equation}
\rho _{G}(\Delta g\pm \widehat{g}(r))=\frac{1}{\pi ^{5}r_{th}^{10}}\exp
\left\{ -\frac{\left( \Delta g\pm \widehat{g}(r)\right) ^{2}}{r_{th}^{2}}%
\right\}  \label{Gaussian PDF at time s}
\end{equation}%
the Gaussian PDF (\ref{REALIZATION OF THE INITIAL ODF}) evaluated for a
generic $4-$position $r(s)$ generally different from the initial one $%
r_{o}\equiv r\left( s_{o}\right) $. Then, it is possible to show that a
formal solution $\rho (G_{L}(s),\Delta g,s)$ of the quantum continuity
equation can more generally be taken of the form%
\begin{equation}
\rho (G_{L}(s),\Delta g,s)=\rho _{G}(\Delta g\pm \widehat{g}(r))\exp \left\{
-\int\limits_{s_{o}}^{s}ds^{\prime }\frac{\partial V_{\mu \nu
}(G_{L}(s^{\prime }),\Delta g,s^{\prime })}{\partial g_{L\mu \nu }(s^{\prime
})}\right\} .  \label{SOLUTION OF CONTINUITY EQ-1}
\end{equation}

Let us display for this purpose an invariance property of the initial PDF.
The following proposition is proved to hold.

\bigskip

\textbf{Proposition \#1 - Invariance of the Gaussian PDF }$\rho _{G}(\Delta
g\pm \widehat{g}(r))$

The following two propositions hold:

P1$_{1})$ \emph{The Gaussian PDF }$\rho _{G}(\Delta g\pm \widehat{g}(r))$%
\emph{\ prescribed by Eq.(\ref{Gaussian PDF at time s}) satisfies the
invariance condition}%
\begin{equation}
\frac{D}{Ds}\ln \rho _{G}(\Delta g\pm \widehat{g}(r))=0.
\label{PROPOSITION-1-C}
\end{equation}

P1$_{2})$ \emph{Eq.(\ref{SOLUTION OF CONTINUITY EQ-1}) realizes a particular
solution of the quantum continuity equation in (\ref{GLP-equations}).}

\emph{Proof - }To prove the invariance property (\ref{PROPOSITION-1-C}) in
proposition P1$_{1}$ one first notices that $\left( \Delta g\pm \widehat{g}%
(r)\right) ^{2}\equiv \left( \Delta g\right) ^{2}\pm 2\Delta g_{\mu \nu }%
\widehat{g}^{\mu \nu }(r)+4$, where%
\begin{equation}
\left\{
\begin{array}{c}
\left( \Delta g(s)\right) ^{2}=\left( \Delta g(s_{o})\right) ^{2}, \\
\Delta g_{\mu \nu }(s)\widehat{g}^{\mu \nu }(r)=\Delta g_{\mu \nu }(s_{o})%
\widehat{g}^{\mu \nu }(r).%
\end{array}%
\right.  \label{PREVIOUS-1}
\end{equation}%
As a consequence it follows that identically $\frac{\mathbf{\ }D}{Ds}\left(
\Delta g(s)\right) ^{2}\equiv 0$,\ while due to the second equation in (\ref%
{PREVIOUS-1})%
\begin{equation}
\frac{D}{Ds}\Delta g_{\mu \nu }(s)\widehat{g}^{\mu \nu }(r)=\Delta g_{\mu
\nu }(s_{o})\frac{D}{Ds}\widehat{g}^{\mu \nu }(r),
\end{equation}%
where one has that identically $\frac{D}{Ds}\widehat{g}^{\mu \nu }(r)\equiv
0 $. Hence, Eq.(\ref{PROPOSITION-1-C}) necessarily holds. This implies in
turn that Eq.(\ref{SOLUTION OF CONTINUITY EQ-1}) is indeed a particular
solution of the quantum continuity equation, as can be easily verified by
algebraic calculation after substitution in the same equation. This proves
proposition P1$_{2}$. \textbf{Q.E.D.}

\subsection{GLP-quantum and stochastic expectation values}

The second implication of Eq.(\ref{PRESCRIPTION OF f(Dg)}) concerns the
prescription of the quantum and stochastic expectation values of arbitrary
observables which are identified with ordinary tensor functions.

Indeed, first, since $\Delta g\equiv \left\{ \Delta g_{\mu \nu }\right\} $
is an observable, $\rho _{o}(\Delta g)$ remains in turn an observable too.
Second, the \emph{quantum expectation values} of quantum observables can be
determined explicitly, \emph{without performing a separate stochastic
average.} In fact, let us consider for definiteness a generic observable
which is represented by an ordinary $s-$dependent real function $X(s)\equiv
X(G_{L}(s),\Delta g,s)$. According to the GLP-representation its quantum
expectation value is given by the configuration-space weighted integral
(hereon referred to as \emph{GLP-quantum expectation value}):%
\begin{equation}
\left\langle X(s)\right\rangle =\int_{U_{g}}d(\delta G_{L})\rho
(G_{L},\Delta g,s)X(G_{L},\Delta g,s),  \label{quantum expect value}
\end{equation}%
where the integration\ is performed with respect to $\delta G_{L}\equiv
\delta G_{L}(s),$\ keeping constant both $\delta g_{L\mu \nu }(s)$\ and the
background metric tensor $\widehat{g}(r)\equiv $\ $\widehat{g}(r(s))$ in
terms of $\rho (G_{L},\Delta g,s)\equiv \rho (G_{L}(s),\Delta g,s)$,\ the
latter being prescribed according to Eq.(\ref{SOLUTION OF CONTINUITY EQ-1}).
One can show that the following equivalent representations of $\left\langle
X(s)\right\rangle $ hold.

\bigskip

\textbf{Proposition \#2 - Equivalent representations of the GLP-quantum
expectation value}\emph{\ }$\left\langle X(s)\right\rangle $

\emph{In validity of Proposition 1 and Eq.(\ref{quantum expect value}) the
following equivalent representations of the GLP-quantum expectation value }$%
\left\langle X(s)\right\rangle $\emph{\ hold:}

\emph{1) First, }$\left\langle X(s)\right\rangle $\emph{\ can be expressed
by means of the expectation value in terms of the initial quantum PDF. This
yields}%
\begin{equation}
\left\langle X(s)\right\rangle =\int_{U_{g}}d(\delta G_{L}(s_{o}))\rho
_{G}(\Delta g\pm \widehat{g}(r))X(G_{L}(s),\Delta g,s),  \label{expect-glp-2}
\end{equation}%
\emph{where the integration is performed on the initial values of the tensor
field }$\delta G_{L}(s_{o})$\emph{\ instead of }$\delta G_{L}(s)$.\emph{\ In
the same integral both }$\delta g_{L\mu \nu }(s)$\emph{\ and }$\widehat{g}%
(r(s))$\emph{\ are again kept constant.}

\emph{2) Second, the same integral can also be equivalently performed in
terms of the integration variable }$\Delta g\equiv \left\{ \Delta g_{\mu \nu
}\right\} $\emph{\ instead of the initial fields }$\delta G_{L}(s_{o}),$%
\emph{\ thus yielding}
\begin{equation}
\left\langle X(s)\right\rangle =\int_{U_{g}}d(\Delta g)\rho _{o}(\Delta g\pm
\widehat{g}(r_{o}))X(G_{L}(s),\Delta g,s)\equiv \left\langle X(s),\widehat{g}%
(r_{o})\right\rangle _{\Delta g},  \label{expect-gkp-3}
\end{equation}%
\emph{where }$\left\langle X(s),\widehat{g}(r_{o})\right\rangle _{\Delta g}$%
\emph{\ identifies the stochastic average of }$X(G_{L}(s),\Delta g,s)$\emph{%
,\ performed in terms of the stochastic PDF }$\rho _{o}(\Delta g\pm \widehat{%
g}(r_{o}))$ \emph{while again keeping constant }$\delta g_{L\mu \nu }(s)$%
\emph{\ and }$\widehat{g}(r(s))$.

\emph{3) Finally, the integral in Eq.(\ref{expect-gkp-3}) can also be
equivalently performed in terms of the integral}
\begin{equation}
\left\langle X(s)\right\rangle =\int_{U_{g}}d(\Delta g)\rho _{o}(\Delta g\pm
\widehat{g}(r(s)))X(G_{L}(s),\Delta g,s)\equiv \left\langle X(s),\widehat{g}%
(r(s))\right\rangle _{\Delta g},  \label{expect-GLP-4}
\end{equation}%
\emph{where }$\left\langle X(s),\widehat{g}(r\left( s\right) )\right\rangle
_{\Delta g}$\emph{\ identifies the stochastic average of }$X(G_{L}(s),\Delta
g,s),$\emph{\ performed in terms of the stochastic PDF }$\rho _{o}(\Delta
g\pm \widehat{g}(r))$ \emph{while keeping constant }$\delta g_{L\mu \nu }(s)$%
\emph{\ and }$\widehat{g}(r)$.

\emph{Proof - }Consider first Eq.(\ref{expect-glp-2}). Its proof follows by
noting that the integral in Eq.(\ref{quantum expect value}) can be
equivalently represented in terms of the inverse mapping $\delta
G_{L}(s)\rightarrow \delta G_{L}(s_{o})$. This implies, in fact, the
differential identity%
\begin{equation}
d(\delta G_{L}(s))=d(\delta G_{L}(s_{o}))\left\vert \frac{\partial \delta
G_{L}(s)}{\partial \delta G_{L}(s_{o})}\right\vert ,  \label{identity-1-bis}
\end{equation}%
where, thanks to Liouville theorem the Jacobian determinant $\left\vert
\frac{\partial \delta G_{L}(s)}{\partial \delta G_{L}(s_{o})}\right\vert $
can be shown to be
\begin{equation}
\left\vert \frac{\partial \delta G_{L}(s)}{\partial \delta G_{L}(s_{o})}%
\right\vert =\exp \left\{ \int\limits_{s_{o}}^{s}ds^{\prime }\frac{\partial
V_{\mu \nu }(G_{L}(s^{\prime }),\Delta g,s^{\prime })}{\partial g_{L\mu \nu
}(s^{\prime })}\right\} .
\end{equation}%
Next, by invoking the solution of the quantum continuity equation (\ref%
{SOLUTION OF CONTINUITY EQ-1}), conservation of probability warrants that%
\begin{equation}
d(\delta G_{L})\rho (G_{L}(s),\Delta g,s)=d(\delta G_{L}(s_{o}))\rho
_{o}(\Delta g\pm \widehat{g}(r(s))),  \label{IDENTITY-1}
\end{equation}%
which in turn implies Eq.(\ref{expect-glp-2}). The proof of Eq.(\ref%
{expect-gkp-3}) is obtained in a similar way by noting that (see Eq.(\ref%
{DSIPALCEMENT 4-TENSOR})) $\Delta g_{\mu \nu }=\delta g_{L\mu \nu
}(s_{o})-\delta G_{L\mu \nu }(s_{o})$ so that the same integral (\ref%
{expect-glp-2}) can also be equivalently performed in terms of the
integration variable $\Delta g\equiv \left\{ \Delta g_{\mu \nu }\right\} $
while keeping constant $\delta g_{L\mu \nu }(s_{o})$ and $\widehat{g}(r)$.
Hence it follows that%
\begin{equation}
d(\delta G_{L}(s_{o}))=d(\Delta g)\left\vert \frac{\partial \delta
G_{L}(s_{o})}{\partial \Delta g}\right\vert =d(\Delta g),
\end{equation}%
since the Jacobian determinant $\left\vert \frac{\partial \delta G_{L}(s_{o})%
}{\partial \Delta g}\right\vert $ is by construction identically equal to $1$%
. Hence the differential identity (\ref{IDENTITY-1}) necessarily holds, thus
yielding also Eq.(\ref{expect-gkp-3}). Finally, the proof of Eq.(\ref%
{expect-GLP-4}) follows from Eq.(\ref{expect-gkp-3}) being an immediate
consequence of Proposition 1. \textbf{Q.E.D.}

\subsection{Generalized Gaussian PDF and emergent gravity interpretation}

Let us examine the implications of the previous Propositions 1 and 2. The
first one concerns the determination of the proper-time evolved quantum PDF $%
\rho (G_{L}(s),\Delta g,s)$, to be based on Proposition 1 (see the
conservation equation (\ref{PROPOSITION-1-C})) and Eq.(\ref{SOLUTION OF
CONTINUITY EQ}). This is given by the equation (\ref{SOLUTION OF CONTINUITY
EQ-1}). Notice that although $\rho _{G}(\Delta g\pm \widehat{g}(r))$\ is a
shifted Gaussian PDF, $\rho (G_{L}(s),\Delta g,s)$\ is generally not so. Its
precise realization depends in fact on the quantum phase-function $%
S^{(q)}(G_{L}(s),\Delta g,s)$, i.e., the corresponding solution of the
quantum H-J equation (in (\ref{GLP-equations})). As a result, the tensor
velocity field $V_{\mu \nu }(G_{L}(s),\Delta g,s)$\ at this stage is still
unknown, thus leaving still undetermined the precise functional form of $%
\rho (G_{L}(s),\Delta g,s),$ so that in general the proper-time evolved PDF $%
\rho (G_{L}(s),\Delta g,s),$\ in contrast to the initial PDF, may be
generally not Gaussian any more. For this reason Eq.(\ref{SOLUTION OF
CONTINUITY EQ-1}) will be referred to in the following as \emph{Generalized
Gaussian PDF}.

The second implication, which is also relevant for the physical
interpretation of the GLP-approach, concerns the following statement.

\bigskip

\textbf{Proposition \#3 - Determination of }$\widehat{g}(r)$\textbf{\
(Emergent gravity)}

\emph{The generalized Gaussian PDF (\ref{SOLUTION OF CONTINUITY EQ-1}) for
all }$r\in \left\{ \mathbf{Q}^{4},\widehat{g}\right\} $ \emph{admits for the
stochastic displacement }$4-$\emph{tensor} $\Delta g_{\mu \nu }$ \emph{the
following GLP-quantum/stochastic expectation value}\textbf{\ }\emph{(in
which both }$\delta g_{L\mu \nu }(s)$\emph{\ and }$\widehat{g}(r(s))$\emph{\
are again kept constant in the integration):}%
\begin{equation}
\left\langle \Delta g_{\mu \nu }\right\rangle \equiv \left\langle \Delta
g_{\mu \nu }\right\rangle _{\Delta g}=\int_{U_{g}}d(\Delta g)\rho
_{G}(\Delta g\pm \widehat{g}(r))\Delta g_{\mu \nu }=\mp \widehat{g}_{\mu \nu
}(r).  \label{EMERGENT GRAVITY}
\end{equation}

\emph{Proof - }The proof follows as an immediate consequence of Proposition
\#2 and in particular thanks to Eq.(\ref{expect-GLP-4}). \textbf{Q.E.D.}

\bigskip

The consequence is that in the whole space-time and for all proper-times $s$%
\ (i.e., for arbitrary $\left( r\equiv r(s),s\right) $) the local value of
the background metric tensor $\widehat{g}(r)$ is prescribed by means of the
GLP-quantum expectation value of the stochastic displacement $4-$tensor\emph{%
\ }$\Delta g_{\mu \nu }$, i.e., $\left\langle \Delta g_{\mu \nu
}\right\rangle $, or equivalently by means of the corresponding stochastic
average $\left\langle \Delta g_{\mu \nu }\right\rangle _{\Delta g}$
evaluated in terms of the stochastic PDF $\rho _{G}(\Delta g\pm \widehat{g}%
(r))$. In this regard one notices that for the validity of Proposition \#3
the initial PDF must be identified with the stochastic PDF $f\left( \Delta g,%
\widehat{g}\right) $, with the latter satisfying the constraint conditions (%
\ref{CONSTRAINTS ON f(Dg)}). This implies the existence of an emergent
gravity phenomenon, in the sense that the background metric tensor $\widehat{%
g}(r)\equiv \widehat{g}(r(s))$\ "emerges" from the quantum gravitational
field $g_{\mu \nu }$ as the quantum/stochastic expectation value of the
stochastic quantum displacement tensor $\Delta g_{\mu \nu }$\ which
characterizes the covariant GLP theory.

The conclusion provides a physical interpretation of CQG-theory.\ Indeed,
consistent with the second-type emergent-gravity paradigm referred to above
(see Introduction), the background space-time appears through a mean-field
gravitational tensor as the result of a suitable ensemble average of an
underlying quantum/stochastic virtual space-time whose quantum-wave dynamics
is described by GLP trajectories. A notable aspect of the conclusion is,
however, that the representation of the proper-time evolved PDF provided by
Eq.(\ref{SOLUTION OF CONTINUITY EQ-1}) is of general character. In fact, Eq.(%
\ref{EMERGENT GRAVITY}) holds independent also of the precise prescription
of the classical/quantum effective potential in the quantum Hamiltonian
operator.\textbf{\ }This means, therefore, that \emph{the emergent-gravity
interpretation of} $\widehat{g}(r)$ \emph{is an intrinsic characteristic
feature of the GLP-representation} developed here for CQG-theory, whereby
the background metric tensor $\widehat{g}(r)$ can be effectively interpreted
as arising from the stochastic fluctuations of GLP trajectories having a
suitable stochastic probability distribution identified with a Gaussian or
more generally Gaussian-like PDF. It follows that $\widehat{g}(r)$ can be
then obtained exactly as a statistical moment in terms of weighted integral
over the stochastic tensor $\Delta g_{\mu \nu }$.

In this sense the concept of emergent gravity proposed here has similarities
with the analogous one to be found in the literature, namely the conjecture
that the geometrical properties of space-time should reveal themselves as a
mean field description of microscopic stochastic or quantum degrees of
freedom underlying the classical solution \cite{emerg1,emerg2}.\textbf{\ }%
However, the physical context proposed here differs from the customary one
adopted in the literature, whereby according to the common emergent gravity
paradigm the Einstein field equations of gravity should have an emergent
character in that, in validity of suitable assumptions, they can be shown to
arise from a thermodynamic approach to space-time \cite{emerg0,emerg3}.

Nevertheless, the explicit construction of particular solutions of the
GLP-parametrized quantum continuity and H-J equations indicated above (see
Eqs.(\ref{GLP-equations})) remains necessary and requires the introduction
of suitable representations both for the quantum phase-function $%
S^{(q)}(G_{L}(s),\Delta g,s)$\ and the quantum effective potential $%
V(G_{L}(s),\Delta g,s)$ (see next Section).

\section{7 - GLP approach: polynomial decomposition of the quantum phase
function}

Based on these premises, we can now implement the GLP formalism and proceed
constructing particular solutions of the quantum H-J equation (see second
equation in (\ref{GLP-equations})).\textbf{\ }More precisely, the goal here
is to look for solutions of the quantum phase function expressed in the
GLP-parametrization, i.e., $S^{(q)}(G_{L}(s),\Delta g,s),$\ which are
expressed by means of polynomial decompositions in terms of power series of
the stochastic tensor $\Delta g$.\textbf{\ }For definiteness, in the sequel
the case is considered in which the following pre-requisites apply:

A) "Harmonic" polynomial decomposition of $S^{(q)}(G_{L}(s),\Delta g,s)$,
i.e., the same quantum phase-function is expressed in terms of a
second-degree polynomial of the form\textbf{\ }%
\begin{equation}
S^{(q)}(G_{L}(s),\Delta g,s)=\frac{a_{pq}^{\alpha \beta }(s)}{2}\Delta
g_{\alpha \beta }\Delta g^{pq}+b_{\alpha \beta }(s)\Delta g^{\alpha \beta
}+c(s),  \label{GLP-POLYNOMIAL REPRESENTATION}
\end{equation}%
with $a_{\mu \nu }^{\alpha \beta }(s),$ $b_{\mu \nu }(s)$ and $c(s)$
denoting respectively suitable real $4-$tensors and a $4-$scalar functions
of $s$ to be determined in terms of the same H-J equation. As shown below,
this implies that the effective kinetic energy $T(G_{L}(s),\Delta g,s)$\
defined by Eq.(\ref{effective kinetic energy}) and\ the Bohm effective
quantum potential $V_{QM}(G_{L}(s),\Delta g,s)$\ prescribed according to Eq.(%
\ref{BOHM-potential}) are both realized by means of polynomials of second
degree in $\Delta g$.

B) An analogous "Harmonic" polynomial decomposition holds for $%
V(G_{L}(s),\Delta g,s)$: namely that a polynomial representation of
analogous type should apply also for the total quantum effective potential
density\ appearing in the quantum H-J equation (see Eq.(\ref{FINAL})).%
\textbf{\ }The latter, to be generally considered of the form $%
V(G_{L}(s),\Delta g,s)$, should therefore admit a polynomial representation
of the type%
\begin{equation}
V(G_{L}(s),\Delta g,s)=\frac{A_{pq}^{\alpha \beta }(s)}{2}\Delta g_{\alpha
\beta }\Delta g^{pq}+B_{\alpha \beta }(s)\Delta g^{\alpha \beta }+C(s),
\label{GLP-POLYNOMIAL REPRESENTATION-2}
\end{equation}%
where the tensor coefficients $A_{\mu \nu }^{\alpha \beta }(s),$ $B_{\mu \nu
}(s)$ and $C(s)$ are considered here functions of $s$ alone to be suitably
determined.

\subsection{Implications of the polynomial \textbf{decomposition} for $%
S^{(q)}(G_{L}(s),\Delta g,s)$}

Let us investigate in detail the consequences of the prescription (\ref%
{GLP-POLYNOMIAL REPRESENTATION}) set on the quantum phase-function $%
S^{(q)}(G_{L}(s),\Delta g,s)$.\textbf{\ }One notices, first, that this
property permits to identify uniquely the proper-time evolved quantum PDF in
terms of a Gaussian PDF, which means that, apart for a proper-time dependent
factor, in such a case\ the PDF $\rho (G_{L}(s),\Delta g,s)$\ becomes
intrinsically non-dispersive in character. In this regard the following
statement holds.

\bigskip

\textbf{Proposition \#4 - Determination of the Gaussian PDF }$\rho
(G_{L}(s),\Delta g,s)$

\emph{In validity of the harmonic polynomial decomposition (\ref%
{GLP-POLYNOMIAL REPRESENTATION}), the generalized Gaussian PDF (\ref%
{SOLUTION OF CONTINUITY EQ-1}) takes the form of the Gaussian PDF}%
\begin{equation}
\rho (G_{L}(s),\Delta g,s)\equiv \rho _{G}(\Delta g+\widehat{g}(r))\exp
\left\{ -16\int\limits_{s_{o}}^{s}dsp^{2}(s^{\prime })\frac{a(s^{\prime })}{%
\alpha L}\right\} ,  \label{STATEMENT}
\end{equation}%
\emph{where }$p(s^{\prime })$\emph{\ and }$a(s^{\prime })$\emph{\ are the }$%
4-$\emph{scalar functions respectively prescribed by Eqs.(\ref{A-4}) and Eq.(%
\ref{DEFINIZIONE DI a(s)}) in Appendix A.}

\emph{Proof - }The proof follows by noting that in this case the tensor
velocity $V_{\mu \nu }(G_{L}(s),\Delta g,s)$\ defined by Eq.(\ref%
{GLP-parametrization of tensor velocity}) becomes explicitly
\begin{equation}
V^{\mu \nu }(G_{L}(s),\Delta g,s)\equiv \frac{1}{\alpha L}\frac{\partial
S^{(q)}(G_{L}(s),\Delta g,s)}{\partial g_{L\mu \nu }(s)}=\frac{%
a_{pq}^{\alpha \beta }(s)}{\alpha L}\frac{\partial \Delta g_{\alpha \beta }}{%
\partial g_{L\mu \nu }(s)}\Delta g^{pq}+\frac{1}{\alpha L}\frac{\partial
\Delta g_{\alpha \beta }}{\partial g_{L\mu \nu }(s)}b^{\alpha \beta }(s).
\label{FIRST-EQ}
\end{equation}%
As a consequence, the divergence of the tensor velocity $\frac{\partial
V^{\mu \nu }(\Delta g,s^{\prime })}{\partial g_{L}^{\mu \nu }(s^{\prime })}$%
, which enters the exponential occurring on the rhs of Eq.(\ref{SOLUTION OF
CONTINUITY EQ}), delivers
\begin{equation}
\frac{\partial V^{\mu \nu }(\Delta g,s^{\prime })}{\partial g_{L}^{\mu \nu
}(s^{\prime })}=\frac{1}{\alpha L}\frac{\partial ^{2}S^{(q)}(\Delta
g,s^{\prime })}{\partial g_{L}^{\mu \nu }(s^{\prime })\partial g_{L\mu \nu
}(s^{\prime })}=\frac{a_{pq}^{\alpha \beta }(s)}{\alpha L}\frac{\partial
\Delta g_{\alpha \beta }}{\partial g_{L\mu \nu }(s^{\prime })}\frac{\partial
\Delta g^{pq}}{\partial g_{L}^{\mu \nu }(s^{\prime })},
\end{equation}%
where the evaluation of the $4-$th order tensor $\frac{\partial \Delta
g_{\alpha \beta }}{\partial g_{L\mu \nu }(s)}$\ is reported in Appendix A
(see, e.g., Eq.(\ref{A-1}) together with Propositions A1 and A2).\textbf{\ }%
Hence the previous equation implies in turn%
\begin{equation}
\frac{\partial V^{\mu \nu }(\Delta g,s)}{\partial g_{L}^{\mu \nu }(s)}%
=p^{2}(s)\frac{a_{pq}^{\alpha \beta }(s)}{\alpha L}\delta _{\alpha \beta
}^{\mu \nu }\delta _{\mu \nu }^{pq}\equiv 16p^{2}(s)\frac{a(s)}{\alpha L},
\end{equation}%
where the notation $\delta _{\alpha \beta }^{\mu \nu }\equiv \delta _{\alpha
}^{\mu }\delta _{\beta }^{\nu }$\ has been introduced. As a consequence, the
proper-time evolved quantum PDF (\ref{SOLUTION OF CONTINUITY EQ-1}) takes
the form (\ref{STATEMENT}). \textbf{Q.E.D.}

\bigskip

Next, let us consider the evaluation of effective kinetic energy $%
T(G_{L}(s),\Delta g,s)$ defined by Eq.(\ref{effective kinetic energy}) and
of the Bohm potential given by Eq.(\ref{BOHM-potential}). Regarding $%
T(G_{L}(s),\Delta g,s),$ thanks again to Eq.(\ref{GLP-POLYNOMIAL
REPRESENTATION}), direct evaluation delivers%
\begin{equation}
T(G_{L}(s),\Delta g,s)=\frac{p^{2}(s)}{2\alpha L}\left[ a_{\mu \nu }^{\alpha
\beta }(s)a_{pq}^{\mu \nu }(s)\Delta g_{\alpha \beta }\Delta g^{pq}+b_{\mu
\nu }(s)b^{\mu \nu }(s)+2a_{\alpha \beta }^{\mu \nu }(s)b_{\mu \nu
}(s)\Delta g^{\alpha \beta }\right] .  \label{FIRST-EQ-EXACT}
\end{equation}%
Concerning instead the Bohm potential, one notices that by invoking
Proposition 4 (i.e., Eq.(\ref{STATEMENT})) the two source terms on the rhs
of Eq.(\ref{BOHM-potential}) become respectively%
\begin{eqnarray}
\frac{\partial \ln \rho (G_{L}(s),\Delta g,s)}{\partial g_{L}^{\mu \nu }(s)}
&=&-\frac{2}{r_{th}^{2}}p(s)\left( \Delta g_{\mu \nu }\pm \widehat{g}_{\mu
\nu }(r)\right) , \\
\frac{\partial ^{2}\ln \rho (G_{L}(s),\Delta g,s)}{\partial g_{L\mu \nu
}(s)\partial g_{L}^{\mu \nu }(s)} &=&-\frac{8}{r_{th}^{2}}p^{2}(s).
\end{eqnarray}%
As a consequence direct substitution in the same equation delivers for the
Bohm potential the representation:%
\begin{eqnarray}
V_{QM}(G_{L}(s),\Delta g,s) &\equiv &-\frac{\hslash ^{2}}{8\alpha L}\left[
\frac{2}{r_{th}^{2}}p(s)\left( \Delta g_{\mu \nu }\pm \widehat{g}_{\mu \nu
}(r)\right) \right] \left[ \frac{2}{r_{th}^{2}}p(s)\left( \Delta g^{\mu \nu
}\pm \widehat{g}^{\mu \nu }(r)\right) \right]  \notag \\
&&-\frac{\hslash ^{2}}{4\alpha L}\left[ -\frac{8}{r_{th}^{2}}p^{2}(s)\right]
,
\end{eqnarray}%
which can be equivalently written as%
\begin{equation}
V_{QM}(G_{L}(s),\Delta g,s)\equiv -\frac{\hslash ^{2}p^{2}(s)}{2\alpha
Lr_{th}^{4}}\left( \Delta g_{\mu \nu }\Delta g^{\mu \nu }\pm 2\widehat{g}%
_{\mu \nu }(r)\Delta g^{\mu \nu }+4\right) +\frac{2\hslash ^{2}p^{2}(s)}{%
\alpha Lr_{th}^{2}}.
\end{equation}

\subsection{Implications of the polynomial decomposition for $%
V(G_{L}(s),\Delta g,s)$}

That an explicit realization of the polynomial representation of the type (%
\ref{GLP-POLYNOMIAL REPRESENTATION-2}) is actually possible for the
effective classical potential density $V(G_{L}(s),\Delta g,s)$ given by Eq.(%
\ref{FINAL}) follows by its definition.\textbf{\ }For definiteness, let us
show how this task can be achieved for a specific realization, i.e., in case
of vacuum. The following proposition holds.

\bigskip

\textbf{Proposition \#5 - Harmonic representation of the vacuum effective
potential}

\emph{The vacuum effective potential (\ref{FINAL}) in the harmonic
polynomial representation (\ref{GLP-POLYNOMIAL REPRESENTATION-2}) takes the
form}%
\begin{equation}
V_{o}(g+\Delta g)=2\sigma \alpha L\Lambda +\sigma \alpha L\Lambda \left[ -%
\frac{1}{2}\Delta g_{\mu \nu }\Delta g^{\mu \nu }-\frac{1}{2}\Delta g^{\mu
\nu }\widehat{g}_{\mu \nu }(r)\Delta g^{\alpha \beta }\widehat{g}_{\alpha
\beta }(r)\right] .  \label{HARMONIC REPRESENTATION OF V_o}
\end{equation}

\emph{Proof - }In fact from Eq.(\ref{VACUUM EFFECTIVE POTENTIAL}) the vacuum
effective potential $V_{o}\left( \widehat{g}+\Delta g\right) $ becomes%
\begin{equation}
V_{o}(g+\Delta g)\equiv \sigma \alpha L\Lambda \left[ 2-\frac{1}{4}\left(
\widehat{g}_{\mu \nu }(r)+\Delta g_{\mu \nu }\right) \left( \widehat{g}^{\mu
\nu }(r)+\Delta g^{\mu \nu }\right) \right] \left[ \left( \widehat{g}%
_{pq}(r)+\Delta g_{pq}\right) \widehat{g}^{pq}(r)-2\right] .
\end{equation}%
The harmonic representation is obtained dropping terms of order $\left(
\Delta g\right) ^{3}$ or higher. When this is done in the previous equation,
Eq.(\ref{HARMONIC REPRESENTATION OF V_o}) is recovered at once. \textbf{%
Q.E.D.}

\bigskip

The form of the source term (\ref{HARMONIC REPRESENTATION OF V_o}) suggests
to seek for the tensor coefficient $a_{\mu \nu }^{\alpha \beta }(s)$ in Eq.(%
\ref{GLP-POLYNOMIAL REPRESENTATION}) a particular realization of the form%
\begin{equation}
a_{pq}^{\alpha \beta }(s)=\frac{1}{2}\left[ a_{(o)}(s)\delta _{pq}^{\alpha
\beta }+a_{(1)}(s)\widehat{g}_{pq}(r)\widehat{g}^{\alpha \beta }(r)\right] ,
\label{SOLUTIPON FOR a-tensotr}
\end{equation}%
so that so that upon invoking Eq.(\ref{DEFINIZIONE DI a(s)}) namely letting,
$a_{\mu \nu }^{\alpha \beta }(s)\delta _{\alpha \beta }^{\mu \nu }\equiv
4a(s)$ it follows $a(s)=\frac{1}{2}\left[ a_{(o)}+a_{(1)}\right] $. As a
consequence, one finds that the tensor coefficients $a_{pq}^{\alpha \beta
}(s)$ in Eq.(\ref{GLP-POLYNOMIAL REPRESENTATION}) can also be written as
\begin{equation}
a_{pq}^{\alpha \beta }(s)=\frac{1}{2}\left[ 2a(s)\delta _{pq}^{\alpha \beta
}+a_{(1)}(s)\left( \widehat{g}_{pq}(r)\widehat{g}^{\alpha \beta }(r)-\delta
_{pq}^{\alpha \beta }\right) \right] .
\end{equation}%
In addition, straightforward algebra yields the identities represented by\
Eqs.(\ref{APP-1-1})-(\ref{APP-1-5}) which are reported in Appendix B.

\subsection{Construction of the GLP-equations}

We now pose the problem of the construction of the set of ODEs which, in
validity of the Harmonic polynomial decompositions indicated above determine
a separable solution of the quantum H-J equation in (\ref{GLP-equations}),
and are thus equivalent to the same equation.\textbf{\ }In the case of the
vacuum effective potential by equating all terms in the polynomial expansion
one obtains a set of ODEs for the $4-$scalar coefficients $a_{(o)}(s),$ $%
a_{(1)}(s)$\ and $c(s)$\ and the $4-$tensor $b_{\alpha \beta }(s)$, here
referred to as \emph{GLP-equations}. These are provided by the first-order
ODEs:%
\begin{equation}
\left\{
\begin{array}{c}
\frac{1}{4}\frac{d}{ds}a_{(o)}(s)=\frac{p^{2}(s)}{8\alpha L}a_{(o)}^{2}(s)-%
\frac{\hslash ^{2}}{2\alpha L}\frac{1}{r_{th}^{4}}p^{2}(s)+\frac{1}{2}\sigma
\alpha L\Lambda +G_{(o)}, \\
\frac{1}{4}\frac{d}{ds}a_{(1)}(s)=\frac{p^{2}(s)}{8\alpha L}\left(
4a_{(1)}^{2}(s)+2a_{(o)}(s)a_{(1)}(s)\right) +\frac{1}{2}\sigma \alpha
L\Lambda +G_{(1)}, \\
\frac{d}{ds}b_{\alpha \beta }(s)=\frac{p^{2}(s)}{2\alpha L}\left[ b_{\alpha
\beta }a_{(o)}(s)+a_{(1)}(s)\widehat{g}_{\alpha \beta }(r)\widehat{g}^{\mu
\nu }(r)b_{\mu \nu }(s)\right] , \\
\frac{d}{ds}c(s)=\frac{p^{2}(s)}{2\alpha L}b_{\mu \nu }(s)b^{\mu \nu }(s)+%
\frac{2\hslash ^{2}}{\alpha L}\frac{1}{r_{th}^{2}}p^{2}(s)+C_{o}(s),%
\end{array}%
\right.   \label{BBB-EQ}
\end{equation}%
where $G_{(o)}$, $G_{(1)}$\ and $C_{o}(s)$\ are in principle arbitrary $4-$%
scalar gauge functions. These can be prescribed in such a way that there it
exists a stationary null solution for the $4-$scalar coefficient $a(s)\equiv
\widehat{a}(s)$,\ namely such that for all $s\in I,$\ $\widehat{a}(s)=0$,\
and hence identically for all $s,s_{o}\in I,$%
\begin{equation}
\left\{
\begin{array}{c}
\widehat{a}_{(o)}(s)\equiv \widehat{a}_{(o)}(s_{o}), \\
\widehat{a}_{(1)}(s)=\widehat{a}_{(1)}(s_{o}), \\
\widehat{a}_{(1)}(s_{o})=-\widehat{a}_{(o)}(s_{o}),%
\end{array}%
\right.   \label{stationary solution}
\end{equation}%
which realizes a particular stationary solution of Eqs.(\ref{BBB-EQ}). This
requires suitably-identifying the gauge functions\ $G_{(o)}$\ and $G_{(1)}$,
which for consistency with Eqs.(\ref{stationary solution}) can always be
prescribed in such a way that%
\begin{equation}
\left\{
\begin{array}{c}
G_{(o)}=-\frac{1}{8\alpha L}\widehat{a}_{(o)}^{2}(s_{o})+\frac{\hslash ^{2}}{%
2\alpha L}\frac{1}{r_{th}^{4}}-\frac{1}{2}\sigma \alpha L\Lambda , \\
G_{(1)}=-\frac{1}{4\alpha L}\widehat{a}_{(o)}^{2}(s_{o})-\frac{1}{2}\sigma
\alpha L\Lambda \equiv 0,%
\end{array}%
\right.
\end{equation}%
so that the first two equations in (\ref{BBB-EQ}) can be written explicitly
as%
\begin{equation}
\left\{
\begin{array}{c}
\frac{1}{4}\frac{d}{ds}a_{(o)}(s)=\frac{1}{8\alpha L}\left[
p^{2}(s)a_{(o)}^{2}(s)-2\alpha ^{2}L^{2}\Lambda \right] -\frac{\hslash ^{2}}{%
2\alpha L}\frac{1}{r_{th}^{4}}\left[ p^{2}(s)-1\right] , \\
\frac{1}{4}\frac{d}{ds}a_{(1)}(s)=\frac{p^{2}(s)}{8\alpha L}\left(
4a_{(1)}^{2}(s)+2a_{(o)}(s)a_{(1)}(s)\right) -\frac{1}{2}\alpha L\Lambda .%
\end{array}%
\right.
\end{equation}%
We finally notice that\ the previous equations can also be conveniently cast
in dimensionless form. Noting that $\left[ a\right] =\left[ a_{(o)}\right] =%
\left[ a_{(1)}\right] =\left[ \hslash \right] =\left[ \alpha \right] $, the
dimensionless representation is obtained by means of the dimensionless
variables
\begin{equation}
\left\{
\begin{array}{c}
\overline{a}_{(o)}(\theta )=\frac{a_{(o)}(\theta )}{\alpha }, \\
\overline{a}_{(1)}(\theta )=\frac{a_{(1)}}{\alpha }, \\
\overline{b}_{\alpha \beta }(\theta )=\frac{b_{\alpha \beta }(s)}{\alpha },
\\
\overline{c}(\theta )=\frac{c}{\alpha }, \\
\theta =\frac{2s}{L}, \\
\overline{\Lambda }=\Lambda L^{2}\cong 9.408,%
\end{array}%
\right.
\end{equation}%
where $\overline{\Lambda }$\ identifies in dimensionless units the
experimental value of cosmological constant, here evaluated in terms of the
Compton Length $L$\ which corresponds to the graviton-mass estimate given in
Ref.\cite{noi6}.\textbf{\ }Then, introducing the notations\textbf{\ }%
\begin{equation}
\left\{
\begin{array}{c}
Y(\theta )\equiv \left( 1+\int\limits_{\theta _{o}}^{\theta }d\theta
^{\prime }\overline{a}(\theta ^{\prime })\right) ^{1/2}, \\
Z\left( \theta \right) =\frac{\overline{a}_{(1)}(\theta )}{Y(\theta )^{2}},%
\end{array}%
\right.
\end{equation}%
equations in (\ref{BBB-EQ}) can be shown to be equivalent to the following
set of ODEs for the coefficients $\overline{a}(\theta )$ and $\overline{a}%
_{(1)}(\theta )$:%
\begin{equation}
\left\{
\begin{array}{c}
\frac{d^{2}}{d\theta ^{2}}Y(\theta )=\frac{3}{16}\frac{Z^{2}(\theta )}{%
Y(\theta )}-\frac{3}{16}\frac{\overline{\Lambda }}{Y(\theta )}+\frac{\hslash
^{2}}{4\alpha ^{2}r_{th}^{4}}\frac{Y(\theta )^{2}-1}{Y(\theta )^{3}}, \\
\frac{d}{d\theta }Z(\theta )=\frac{Z^{2}(\theta )}{Y(\theta )}-\frac{1}{2}%
\frac{\overline{\Lambda }}{Y(\theta )},%
\end{array}%
\right.   \label{GLP-equations-1}
\end{equation}%
which admit the stationary solution%
\begin{equation}
\left\{
\begin{array}{c}
\overline{a}(s)=0, \\
\overline{a}_{(o)}^{2}(s)-\frac{1}{2}\overline{\Lambda }=0.%
\end{array}%
\right.   \label{STATIONARY DIMENSONLESS SOLUTION}
\end{equation}

\subsection{Small-amplitude solutions - Conditions of validity}

Now we look for small-amplitude solutions of Eqs.(\ref{GLP-equations-1}).
For definiteness let us introduce the representations%
\begin{equation}
\left\{
\begin{array}{c}
Y(\theta )=Y(\theta _{o})+\delta Y(\theta ), \\
Z(\theta )=Z(\theta _{o})+\delta Z(\theta ),%
\end{array}%
\right.
\end{equation}%
with $Y(\theta _{o})=1,$ $Z(\theta _{o})=\overline{a}_{(1)}(\theta _{o})=\pm
\sqrt{\frac{1}{2}\overline{\Lambda }}$\textbf{\ }and $\delta Y(\theta ),$\ $%
\delta Z(\theta )$\ denoting displacements such that for all\textbf{\ }$%
\theta \in I_{s_{\theta o}}^{\left( +\right) }\equiv \left[ \theta
_{o},+\infty \right] $%
\begin{eqnarray}
0 &<&\delta Y(\theta )\ll 1,  \notag \\
0 &<&\left\vert \delta Z(\theta )/\sqrt{\frac{1}{2}\overline{\Lambda }}%
\right\vert \ll 1.  \label{ORDERINGS}
\end{eqnarray}%
These will be denoted as small-amplitude\ solutions. In this regard the
following proposition holds.

\bigskip

\textbf{Proposition \#6 - Small-amplitude solutions of Eqs.(\ref%
{GLP-equations-1})}

\emph{For all }$s\in I_{s_{o}}^{\left( +\right) }\equiv \left[ s_{o},+\infty %
\right] $\emph{\ Eqs.(\ref{GLP-equations-1}) admit small-amplitude solutions.%
}

\emph{Proof - }In fact upon linearization Eqs.(\ref{GLP-equations-1}) imply
respectively%
\begin{equation}
\left\{
\begin{array}{c}
\frac{d^{2}}{d\theta ^{2}}\delta Y(\theta )=\frac{3}{16}\left[ \pm 2\sqrt{%
\frac{1}{2}\overline{\Lambda }}\delta Z-\frac{1}{2}\overline{\Lambda }\delta
Y(\theta )\right] +\frac{3}{16}\overline{\Lambda }\delta Y(\theta )+\frac{%
\hslash ^{2}}{4\alpha ^{2}r_{th}^{4}}2\delta Y(\theta ), \\
\frac{d}{d\theta }\delta Z(\theta )=\pm 2\sqrt{\frac{1}{2}\overline{\Lambda }%
}\delta Z.%
\end{array}%
\right.
\end{equation}%
The two equations deliver respectively the solutions%
\begin{equation}
\left\{
\begin{array}{c}
\delta Z(\theta )=\delta Z(\theta _{o})\exp \left\{ \pm 2\sqrt{\frac{1}{2}%
\overline{\Lambda }}(\theta -\theta _{o})\right\} , \\
\delta a(\theta )=A\delta Z(\theta _{o})\exp \left\{ \pm 2\sqrt{\frac{1}{2}%
\overline{\Lambda }}(\theta -\theta _{o})\right\} ,%
\end{array}%
\right.  \label{SMALL-AMP-SOL}
\end{equation}%
with $A$\ denoting the constant coefficient
\begin{equation}
A=\frac{\frac{3}{4}\overline{\Lambda }}{1-\frac{9}{32}\overline{\Lambda }-%
\frac{\hslash ^{2}}{2\alpha ^{2}r_{th}^{4}}}.
\end{equation}%
As a consequence, Eqs.(\ref{SMALL-AMP-SOL}) imply also that%
\begin{equation}
\delta a_{(1)}(\theta )=\delta Z(\theta )\left[ 1+\frac{A}{2}\right] .
\label{STAR-2}
\end{equation}%
\textbf{Q.E.D.}

\bigskip

Thus, we conclude that small-amplitude solutions of the GLP-equations (\ref%
{GLP-equations-1}) indeed exist which depend exponentially on proper time,
the exponential factor being of the form $\exp \left\{ -2\sqrt{\frac{1}{2}%
\overline{\Lambda }}(\theta -\theta _{o})\right\} $\ or $\exp \left\{ 2\sqrt{%
\frac{1}{2}\overline{\Lambda }}(\theta -\theta _{o})\right\} $ respectively.
These are referred to respectively as \emph{decay }and \emph{blow-up}
small-amplitude solutions. In the two cases for $\theta -\theta _{o}$\ $%
\rightarrow +\infty $,\ these either decay to the constant solution or
diverge exponentially.\ Therefore, quantum stationary solutions can be
identified with asymptotic ones, i.e., as final states of decaying quantum
solutions. Blow-up solutions, however, for finite times $\theta -\theta _{o}$
$>0$ necessarily\ violate the ordering assumptions (\ref{ORDERINGS}) and as
such Eqs.(\ref{SMALL-AMP-SOL}) and (\ref{STAR-2}) are no more applicable in
such a case.

The investigation of the blow-up solutions requires therefore the proper
consideration of the set of GLP-equations (\ref{GLP-equations-1}). One can
show, however, that if the following asymptotic orderings apply%
\begin{eqnarray}
Y(\theta ) &\gg &1, \\
\left\vert Z(\theta )/\sqrt{\frac{1}{2}\overline{\Lambda }}\right\vert &\gg
&1,
\end{eqnarray}%
then in such a case the asymptotic limits must apply%
\begin{eqnarray}
\lim_{\theta -\theta _{o}\rightarrow +\infty }Y(\theta ) &=&+\infty , \\
\lim_{\theta -\theta _{o}\rightarrow +\infty }\frac{d}{d\theta }Y(\theta )
&=&0.
\end{eqnarray}%
These imply in turn also the vanishing of the $4-$scalar coefficient $p(s)$\
(see Appendix A) in the proper-time limit $s-s_{o}\rightarrow +\infty $,
i.e.,%
\begin{equation}
\lim_{s-s_{o}\rightarrow +\infty }p(s)=0.  \label{LIMIT-3}
\end{equation}%
The implication of Eq.(\ref{LIMIT-3}) is however the violation in the same
limit of the Heisenberg inequality%
\begin{equation}
\left\langle \left( \Delta g_{_{(\mu )(\nu )}}\right) ^{2}\right\rangle
\left\langle \left( \Delta \pi _{\mu \nu }\right) ^{2}\right\rangle _{1}\geq
\frac{\hbar ^{2}}{4},  \label{HEISENBERG}
\end{equation}%
pointed out in Ref.\cite{noi7}, with $\left\langle \left( \Delta g_{_{(\mu
)(\nu )}}\right) ^{2}\right\rangle $ and $\left\langle \left( \Delta \pi
_{\mu \nu }\right) ^{2}\right\rangle _{1}$ denoting respectively%
\begin{eqnarray}
&&\left. \left\langle \left( \Delta \pi _{_{\mu \nu }}\right)
^{2}\right\rangle _{1}=\frac{\hbar ^{2}}{4}\int_{U_{g}}d(g)\rho \frac{%
\partial \ln \rho }{\partial g^{\mu \nu }}\frac{\partial \ln \rho }{\partial
g^{(\mu )(\nu )}},\right. \\
\left\langle \left( \Delta g_{\mu \nu }\right) ^{2}\right\rangle
&=&\int_{U_{g}}d(g)\rho \left( g_{\mu \nu }-\widetilde{g}_{\mu \nu }\right)
\left( g_{(\mu )(\nu )}-\widetilde{g}_{(\mu )(\nu )}\right) =\frac{1}{10}%
r_{th}^{2}.
\end{eqnarray}%
In fact, due to Eq.(\ref{LIMIT-3}) it follows that
\begin{equation}
\lim_{s-s_{o}\rightarrow +\infty }\left\langle \left( \Delta g_{\mu \nu
}\right) ^{2}\right\rangle =0.
\end{equation}

Instead, one can show that constant or small-amplitude decaying solutions
satisfy the Heisenberg inequality (\ref{HEISENBERG}) and as such realize
physically admissible quantum solutions. Such a conclusion, therefore, rules
out blow-up solutions from the class of physically-admissible solutions in
the same limit.

\section{8 - Conclusions}

In this paper the basic principles\ of a new trajectory-based approach to
manifestly-covariant quantum gravity (CQG) theory have been laid down. This
provides new physical insight into the nature and behavior of the
manifestly-covariant quantum-wave equation and corresponding equivalent set
of quantum hydrodynamic equations that are realized by means of CQG-theory.
For its similarity with the analogous Generalized Lagrangian Path approach
holding in non relativistic quantum mechanics \cite{Tessarotto2016}, this is
referred to here as Generalized Lagrangian Path (GLP) approach (or
representation) of CQG-theory.

The GLP approach presented here has been shown to be ontologically
equivalent to the "standard" formulation of CQG-theory based on the Eulerian
CQG-wave equation. This occurs because, provided the stochastic PDF $%
f(\Delta g,\widehat{g})$ is identified with the Gaussian PDF $\rho
_{G}(\Delta g\pm \widehat{g}(r_{o}))$ defined above (see Eq.(\ref{Gaussian
PDF at time s})), it does not require any\ kind of addition/modification of
the related fundamental axioms established in Ref.\cite{noi6}.\textbf{\ }%
This feature permits one to effectively reconcile the Eulerian and
Lagrangian descriptions of covariant quantum gravity, which are achieved
respectively in terms of the Eulerian and GLP representations of CQG-wave
equation and of the quantum wave-function. Nevertheless, it\ also provides a
statistical generalization of the Bohmian interpretation of quantum gravity
based on the notion of unique, i.e., deterministic, configuration-space
Lagrangian trajectories belonging to the configuration space $U_{g}$ spanned
by the symmetric tensor field $g\equiv \left\{ g_{\mu \nu }\right\} $. In
fact, in the framework of GLP-theory, each Bohmian trajectory is associated
with an infinite ensemble of stochastic Lagrangian trajectories associated
with the stochastic tensor variable $\Delta g_{\mu \nu }$. Thus, GLP
trajectories replace the customary deterministic Lagrangian trajectories
(LPs) adopted in the original Bohmian approach, from which they inherently
differ for their stochastic character. As a consequence, it is shown that it
is possible to replace each LP with a corresponding continuum set of
stochastic GLP.

A further notable aspect of the GLP approach is, however, that it realizes
at the same time also a solution method for the CQG-wave equation and the
corresponding equivalent quantum hydrodynamic equations. This is obtained by
means of the explicit parametrization of the same equations (and of the
quantum wave-function) in terms of the stochastic displacement tensor $%
\Delta g_{\mu \nu }$ introduced here (see Eq.(\ref{SETTING})). As an
application of the theory developed in this paper, the problem of
constructing\ Gaussian or Gaussian-like solutions of the CQG-wave equation
has been addressed. For this purpose, the case of vacuum fields, i.e.,
obtained in the absence of external classical sources but with the inclusion
of a non-vanishing cosmological constant, has been considered. In this
connection the explicit construction of solutions of the CQG-quantum
hydrodynamic equations has been carried out in which the GLP-parametrized
quantum wave function $\psi (G_{L}(s),\Delta g,s)$ is characterized by a
globally-defined Gaussian-like or Gaussian PDF which satisfies identically
the corresponding quantum continuity equation.\textbf{\ }As a notable
result, the validity of the emergent-gravity picture has been demonstrated,
referred to here as\textbf{\ }"\textit{second-type emergent-gravity\ paradigm%
}".\textbf{\ }Accordingly, the background space-time metric tensor\ $%
\widehat{g}_{\mu \nu }(r)$\ of CQG-theory has been identified in terms of a
suitable quantum/stochastic expectation value of the quantum state, i.e.,
weighted in terms of the corresponding quantum PDF.

In addition, the problem of the construction of separable solutions of the
quantum Hamilton-Jacobi (H-J) equation has been posed which satisfy at the
same time also the requirements that the quantum wave function $\psi
(G_{L}(s),\Delta g,s)$\ is dynamically consistent, in the sense that the
corresponding (GLP-parametrized) quantum PDF $\rho (G_{L}(s),\Delta g,s)$\
associated with the quantum wave-function is globally conserved. The
solution of the H-J equation has been based on the polynomial
representations of the quantum effective potential. In particular, separable
solutions for the GLP-parametrized quantum phase function $S(G_{L}(s),\Delta
g,s)$\ have been determined based on a harmonic (i.e., second degree)
polynomial expansion with respect to the stochastic displacement tensor $%
\Delta g_{\mu \nu }$.\ The coefficients of the same expansion have been
shown to satisfy an equivalent set of first-order evolution ODEs, denoted as
GLP-equations. The same coefficients admit both stationary and
non-stationary solutions with respect to the dependence on the background
proper-time $s$.\textbf{\ }Non-stationary solutions include, in particular,
the case of small-amplitude solutions which remain globally (i.e., for all $%
s $\ greater than the initial proper-time $s_{o}$) suitably close to the
stationary ones. These have been identified here with particular solutions
exponentially decaying (to the constant ones).

These conclusions show that particular solutions of the\ CQG-quantum
wave-equation exist which are characterized by Gaussian quantum PDF.
Remarkably, the same solutions can be either stationary, i.e., characterized
by quantum wave-functions of the type $\psi =\psi (G_{L}(s),\Delta g),$\ or
non-stationary ones $\psi (G_{L}(s),\Delta g,s)$, namely depending
explicitly on the proper-time $s$.\textbf{\ }This scenario is promising for
its possible implications suggesting that the investigation of
non-stationary solutions of the quantum wave-function may be actually an
important and challenging subject of future research in quantum gravity,
quantum cosmology and CQG-theory.

\begin{quote}
\textit{Acknowledgments - }Work developed within the research project of the
Albert Einstein Center for Gravitation and Astrophysics, Czech Science
Foundation No. 14-37086G (M.T.). The publication of this work was supported
by the Fetzer Franklin Fund of the John E. Fetzer Memorial Trust.

\bigskip
\end{quote}

\section{Appendix A - Evaluation of $p(s)$ and differential identities}

In this appendix the proof of Eq.(\ref{STATEMENT}) in Proposition \#4 and
the determination of the $4-$scalar factor $p(s)$ are explicitly pointed out
in the following propositions.

\bigskip

\textbf{Proposition A1 - Determination of the tensor field }$\frac{\partial
\Delta g_{\alpha \beta }}{\partial g_{L\mu \nu }(s^{\prime })}$

\emph{Given validity of the polynomial representation (\ref{GLP-POLYNOMIAL
REPRESENTATION}), the tensor field }$\frac{\partial \Delta g_{\alpha \beta }%
}{\partial g_{L\mu \nu }(s^{\prime })}$\emph{\ takes the form}%
\begin{equation}
\frac{\partial \Delta g_{\mu ^{\prime }\nu ^{\prime }}}{\partial g_{L\mu \nu
}(s^{\prime })}=-\frac{\partial \Delta g_{\mu ^{\prime }\nu ^{\prime }}}{%
\partial G_{L\mu \nu }(s^{\prime })},  \label{A-0}
\end{equation}%
\emph{with}%
\begin{equation}
\frac{\partial \Delta g_{\mu ^{\prime }\nu ^{\prime }}}{\partial g_{L\mu \nu
}(s^{\prime })}=\delta _{\mu ^{\prime }}^{\mu }\delta _{\nu ^{\prime }}^{\nu
}p(s),  \label{A-1}
\end{equation}%
\emph{and }$p(s)$\emph{\ is the }$4-$\emph{scalar function determined by the
integral equation}%
\begin{equation}
p(s)=\frac{1}{1+\int_{s_{o}}^{s}ds^{\prime }\frac{1}{\alpha L}a(s^{\prime
})g(s^{\prime })}.  \label{A-1ter}
\end{equation}%
\emph{Here }$a(s)\equiv \frac{1}{16}a_{\alpha \beta }^{pq}(s)\delta
_{pq}^{\alpha \beta }$\emph{\ and }$a_{\alpha \beta }^{pq}(s)$\emph{\ is the
tensor introduced in the polynomial decomposition of the phase function }$%
S^{(q)}$\emph{\ given by Eq.(\ref{GLP-POLYNOMIAL REPRESENTATION}).}

\emph{Proof - }One first notices that, provided the quantum phase function
is of the form $S^{(q)}=S^{(q)}(\Delta g,s^{\prime })$,\ and noting that $%
\delta g_{L\mu \nu }^{(o)}=\delta G_{L\mu \nu }^{(o)}+\Delta g_{\mu \nu }$,
then the LP-initial-value problem (\ref{LP-3}) delivers
\begin{equation}
\delta g_{L\mu \nu }(s)=\delta g_{L\mu \nu
}^{(o)}+\int_{s_{o}}^{s}ds^{\prime }\frac{1}{\alpha L}\frac{\partial
S^{(q)}(\Delta g,s^{\prime })}{\partial g_{L}^{\mu \nu }(s^{\prime })},
\end{equation}%
or equivalently%
\begin{equation}
\delta g_{L\mu \nu }(s)=\delta G_{L\mu \nu }^{(o)}+\Delta g_{\mu \nu
}+\int_{s_{o}}^{s}ds^{\prime }\frac{1}{\alpha L}\frac{\partial
S^{(q)}(\Delta g,s^{\prime })}{\partial g_{L}^{\mu \nu }(s^{\prime })}.
\label{NEW A-1}
\end{equation}%
The last equation therefore implies also that the solution to the
GLP-initial-value problem (\ref{GLP-initial-value problem}) is similarly%
\begin{equation}
\delta G_{L\mu \nu }(s)=\delta g_{L\mu \nu }(s_{o})-\Delta g_{\mu \nu
}+\int_{s_{o}}^{s}ds^{\prime }\frac{1}{\alpha L}\frac{\partial
S^{(q)}(\Delta g,s^{\prime })}{\partial g_{L}^{\mu \nu }(s^{\prime })}.
\label{NEW A-2}
\end{equation}%
Then, differentiating Eq.(\ref{NEW A-1}) with respect to $\delta g_{L\mu \nu
}(s)$ while keeping $\delta G_{L\mu \nu }(s_{o})\equiv \delta G_{L\mu \nu
}^{(o)}$ constant, yields%
\begin{equation}
\delta _{\mu ^{\prime }}^{\mu }\delta _{\nu ^{\prime }}^{\nu }\equiv \frac{%
\partial g_{L\mu ^{\prime }\nu ^{\prime }}(s)}{\partial g_{L\mu \nu }(s)}=%
\frac{\partial \Delta g_{\mu ^{\prime }\nu ^{\prime }}}{\partial g_{L\mu \nu
}(s)}+\frac{\partial \Delta g_{\alpha \beta }}{\partial g_{L\mu \nu }(s)}%
\frac{\partial }{\partial \Delta g_{\alpha \beta }}\int_{s_{o}}^{s}ds^{%
\prime }\frac{1}{\alpha L}\frac{\partial S^{(q)}(\Delta g,s^{\prime })}{%
\partial g_{L}^{\mu ^{\prime }\nu ^{\prime }}(s^{\prime })},
\label{PRIMO-CASO}
\end{equation}%
where in the following we shall adopt the short notation $\delta _{\mu
^{\prime }\nu ^{\prime }}^{\mu \nu }\equiv \delta _{\mu ^{\prime }}^{\mu
}\delta _{\nu ^{\prime }}^{\nu }$ and by construction%
\begin{equation}
\frac{\partial S^{(q)}(\Delta g,s^{\prime })}{\partial G_{L}^{\mu ^{\prime
}\nu ^{\prime }}(s^{\prime })}=-\frac{\partial S^{(q)}(\Delta g,s^{\prime })%
}{\partial g_{L}^{\mu ^{\prime }\nu ^{\prime }}(s^{\prime })},
\end{equation}%
and hence%
\begin{equation}
\frac{\partial \Delta g_{\mu ^{\prime }\nu ^{\prime }}}{\partial G_{L}^{\mu
\nu }(s)}=-\frac{\partial \Delta g_{\mu ^{\prime }\nu ^{\prime }}}{\partial
g_{L}^{\mu \nu }(s)}.
\end{equation}%
As a consequence, if one performs the differentiation of Eq.(\ref{NEW A-2})
with respect to $G_{L\mu \nu }(s)$ while keeping $\delta g_{L\mu \nu
}(s_{o})\equiv \delta g_{L\mu \nu }^{(o)}$ as constant,\textbf{\ }it follows
equivalently that%
\begin{equation}
\delta _{\mu ^{\prime }}^{\mu }\delta _{\nu ^{\prime }}^{\nu }\equiv \frac{%
\partial G_{L\mu ^{\prime }\nu ^{\prime }}(s)}{\partial G_{L\mu \nu }(s)}=-%
\frac{\partial \Delta g_{\mu ^{\prime }\nu ^{\prime }}}{\partial G_{L\mu \nu
}(s)}-\frac{\partial \Delta g_{\alpha \beta }}{\partial G_{L\mu \nu }(s)}%
\frac{\partial }{\partial \Delta g_{\alpha \beta }}\int_{s_{o}}^{s}ds^{%
\prime }\frac{1}{\alpha L}\frac{\partial S^{(q)}(\Delta g,s^{\prime })}{%
\partial g_{L}^{\mu ^{\prime }\nu ^{\prime }}(s^{\prime })}.  \label{A-2}
\end{equation}%
Therefore, from Eq.(\ref{PRIMO-CASO}) denoting $\delta _{\mu ^{\prime }\nu
^{\prime }}^{\mu \nu }\equiv \delta _{\mu ^{\prime }}^{\mu }\delta _{\nu
^{\prime }}^{\nu }$ it follows%
\begin{equation}
\delta _{\mu \nu }^{\mu ^{\prime }\nu ^{\prime }}=\frac{\partial \Delta
g^{\mu ^{\prime }\nu ^{\prime }}}{\partial g_{L}^{\mu \nu }(s)}+\frac{%
\partial \Delta g^{\alpha \beta }}{\partial g_{L}^{\mu \nu }(s)}\frac{%
\partial }{\partial \Delta g^{\alpha \beta }}\int_{s_{o}}^{s}ds^{\prime }%
\frac{1}{\alpha L}\frac{\partial S^{(q)}(\Delta g,s^{\prime })}{\partial
g_{L\mu ^{\prime }\nu ^{\prime }}(s^{\prime })},  \label{A-3}
\end{equation}%
where due to the polynomial representation (\ref{GLP-POLYNOMIAL
REPRESENTATION})%
\begin{eqnarray}
&&\left. \frac{\partial S^{(q)}(g_{L}(s^{\prime }),\Delta g,s^{\prime })}{%
\partial g_{L\mu ^{\prime }\nu ^{\prime }}(s^{\prime })}=\frac{\partial
\Delta g_{pq}}{\partial g_{L\mu ^{\prime }\nu ^{\prime }}(s^{\prime })}\left[
a_{p^{\prime }q^{\prime }}^{pq}(s^{\prime })\Delta g^{p^{\prime }q^{\prime
}}+b^{pq}(s)\right] \right. , \\
&&\left. \frac{\partial }{\partial \Delta g^{\alpha \beta }}\frac{\partial
S^{(q)}(g_{L}(s^{\prime }),\Delta g,s^{\prime })}{\partial g_{L\mu ^{\prime
}\nu ^{\prime }}(s^{\prime })}=a_{\alpha \beta }^{\mu \nu }(s^{\prime })%
\frac{\partial \Delta g_{\mu \nu }}{\partial g_{L\mu ^{\prime }\nu ^{\prime
}}(s^{\prime })}\right. .
\end{eqnarray}%
As a result Eq.(\ref{A-3}) delivers%
\begin{equation}
\delta _{\mu \nu }^{\mu ^{\prime }\nu ^{\prime }}=\frac{\partial \Delta
g^{\mu ^{\prime }\nu ^{\prime }}}{\partial g_{L}^{\mu \nu }(s)}+\frac{%
\partial \Delta g^{\alpha \beta }}{\partial g_{L}^{\mu \nu }(s)}%
\int_{s_{o}}^{s}ds^{\prime }\frac{a_{\alpha \beta }^{pq}(s^{\prime })}{%
\alpha L}\frac{\partial \Delta g_{pq}}{\partial g_{L\mu ^{\prime }\nu
^{\prime }}(s^{\prime })},
\end{equation}%
thus implying validity of Eq.(\ref{A-1}). In fact, thanks to Eq.(\ref{A-1})
we can write the previous equation as%
\begin{equation}
\delta _{\mu \nu }^{\mu ^{\prime }\nu ^{\prime }}=\delta _{\mu \nu }^{\mu
^{\prime }\nu ^{\prime }}g\left( s\right) +\delta _{\mu \nu }^{\alpha \beta
}g\left( s\right) \int_{s_{o}}^{s}ds^{\prime }\frac{a_{\alpha \beta
}^{pq}(s^{\prime })}{\alpha L}\delta _{pq}^{\mu ^{\prime }\nu ^{\prime
}}g\left( s^{\prime }\right) .
\end{equation}%
Then, defining%
\begin{equation}
a\left( s^{\prime }\right) \delta _{\mu \nu }^{\mu ^{\prime }\nu ^{\prime
}}\equiv \delta _{\mu \nu }^{\alpha \beta }a_{\alpha \beta }^{pq}(s^{\prime
})\delta _{pq}^{\mu ^{\prime }\nu ^{\prime }}  \label{DEFINITION OF a(s)}
\end{equation}%
and substituting, after simplification we get that%
\begin{equation}
g(s)\left[ 1+\int_{s_{o}}^{s}ds^{\prime }\frac{1}{\alpha L}a(s^{\prime
})g(s^{\prime })\right] =1,  \label{A-1bis}
\end{equation}%
while straightforward algebra yields%
\begin{equation}
a(s^{\prime })=\frac{1}{16}\delta _{\mu \nu }^{\alpha \beta }a_{\alpha \beta
}^{pq}(s^{\prime })\delta _{pq}^{\mu ^{\prime }\nu ^{\prime }}\delta _{\mu
^{\prime }\nu ^{\prime }}^{\mu \nu }\equiv \frac{1}{16}a_{\alpha \beta
}^{pq}(s^{\prime })\delta _{pq}^{\alpha \beta }.  \label{DEFINIZIONE DI a(s)}
\end{equation}%
Thus provided $1+\int_{s_{o}}^{s}ds^{\prime }\frac{1}{\alpha L}a(s^{\prime
})g(s^{\prime })\neq 0,$ Eq.(\ref{A-1ter}) follows. \textbf{Q.E.D.}

\bigskip

\textbf{Proposition A2 - Determination of the }$4-$\textbf{scalar function }$%
p(s)$

\emph{In validity of Eq.(\ref{A-1ter}) it follows that}%
\begin{equation}
\left\vert p(s)\right\vert =\frac{1}{\left( 1+\frac{2}{\alpha L}%
\int\limits_{s_{o}}^{s}ds^{\prime }a(s^{\prime })\right) ^{1/2}}.
\label{A-4}
\end{equation}%
\emph{Proof - }In fact, if $p(s)\neq 0$ Eq.(\ref{A-1bis}) implies%
\begin{equation}
1+\int_{s_{o}}^{s}ds^{\prime }\frac{1}{\alpha L}a(s^{\prime })g(s^{\prime })=%
\frac{1}{p(s)}.
\end{equation}%
Differentiating the same equation term by term with respect to $s$ yields
the ODE%
\begin{equation}
\frac{1}{\alpha L}a(s)p(s)=-\frac{p^{\prime }(s)}{p^{2}(s)}.
\end{equation}%
This can be solved noting that $p(s_{o})=1$. Thus one finds%
\begin{equation}
\frac{1}{2p\left( s\right) ^{2}}-\frac{1}{2}=\frac{1}{\alpha L}%
\int\limits_{s_{o}}^{s}ds^{\prime }a(s^{\prime }),
\end{equation}%
whose solution is given by Eq.(\ref{A-4}). \textbf{Q.E.D.}

\section{Appendix B - Differential identities for \textbf{the tensor
coefficients }$a_{pq}^{\protect\alpha \protect\beta }(s)$}

In this appendix the explicit calculations are reported of a number of
useful identities invoked in Section 7. First, one notices that invoking Eq.(%
\ref{SOLUTIPON FOR a-tensotr}) it follows that%
\begin{equation}
a_{\mu \nu }^{\alpha \beta }(s)a_{pq}^{\mu \nu }(s)=\frac{1}{4}\left[
a_{(o)}^{2}(s)\delta _{pq}^{\alpha \beta }+\left(
4a_{(1)}^{2}(s)+2a_{(o)}(s)a_{(1)}(s)\right) \widehat{g}_{pq}(r)\widehat{g}%
^{\alpha \beta }(r)\right] ,  \label{APP-1-1}
\end{equation}%
and similarly%
\begin{equation}
4a_{(1)}^{2}(s)+2a_{(o)}(s)a_{(1)}(s)=2a_{(1)}^{2}(s)+4a(s)a_{(1)}(s),
\label{APP-1-6}
\end{equation}%
\begin{eqnarray}
&&\left. a_{(o)}^{2}(s)+4a_{(1)}^{2}(s)+2a_{(o)}(s)a_{(1)}(s)=\right.
\notag \\
&&\left. =\left[ a_{(o)}(s)+a_{(1)}(s)\right]
^{2}+3a_{(1)}^{2}(s)=4a^{2}(s)+3a_{(1)}^{2}(s).\right.   \label{APP-1-7}
\end{eqnarray}%
The prescription (\ref{SOLUTIPON FOR a-tensotr}) implies therefore:

A) from Eq.(\ref{GLP-POLYNOMIAL REPRESENTATION-2}):%
\begin{eqnarray}
\frac{D}{Ds}S^{(q)}(\Delta g,s) &=&\frac{1}{4}\Delta g_{\alpha \beta }\Delta
g^{\mu \nu }\left[ \frac{d}{ds}a_{(o)}(s)\delta _{\mu \nu }^{\alpha \beta }+%
\widehat{g}_{\mu \nu }(r)\widehat{g}^{\alpha \beta }(r)\frac{d}{ds}a_{(1)}(s)%
\right]   \notag \\
&&+\Delta g^{\alpha \beta }\frac{d}{ds}b_{\alpha \beta }(s)+\frac{d}{ds}c(s);
\label{APP-1-2}
\end{eqnarray}

B) from Eq.(\ref{FIRST-EQ-EXACT}):%
\begin{equation}
\frac{\partial S^{(q)}(G_{L}(s),\Delta g,s)}{\partial g_{L}^{\mu \nu }(s)}=%
\frac{1}{2}\left[ a_{(o)}(s)\delta _{\mu \nu }^{\alpha \beta }+a_{(1)}(s)%
\widehat{g}_{\mu \nu }(r)\widehat{g}^{\alpha \beta }(r)\right] \Delta
g_{\alpha \beta }p(s)+b_{\mu \nu }(s)p(s).  \label{AAP-1-3}
\end{equation}%
Hence, the quantum $4-$tensor fluid velocity field can be represented as%
\begin{equation}
V_{\mu \nu }=\frac{1}{2\alpha L}\left[ a_{(o)}(s)\delta _{\mu \nu }^{\alpha
\beta }+a_{(1)}(s)\widehat{g}_{\mu \nu }(r)\widehat{g}^{\alpha \beta }(r)%
\right] \Delta g_{\alpha \beta }p(s)+\frac{1}{\alpha L}b_{\mu \nu }(s)p(s),
\label{APP-1-4-tensor fluid velocity}
\end{equation}%
with\ the first term on the rhs, linearly proportional to $\Delta g,$\
representing the stochastic part of the quantum fluid velocity.\textbf{\ }%
Similarly one obtains that in Eq.(\ref{FIRST-EQ-EXACT}) the following
identities hold:%
\begin{equation}
\left\{
\begin{array}{c}
a_{\mu \nu }^{\alpha \beta }(s)a_{pq}^{\mu \nu }(s)=\frac{1}{4}\left[
a_{(o)}^{2}(s)\delta _{pq}^{\alpha \beta }+\left(
4a_{(1)}^{2}(s)+2a_{(o)}(s)a_{(1)}(s)\right) \widehat{g}_{pq}(r)\widehat{g}%
^{\alpha \beta }(r)\right] , \\
2a_{\alpha \beta }^{\mu \nu }(s)b_{\mu \nu }(s)=\left[ a_{(o)}(s)\delta
_{\mu \nu }^{\alpha \beta }+a_{(1)}(s)\widehat{g}_{\mu \nu }(r)\widehat{g}%
^{\alpha \beta }(s)\right] b^{\mu \nu }(s).%
\end{array}%
\right.   \label{APP-1-5}
\end{equation}

\end{document}